\documentclass[11pt]{article}

\usepackage{amssymb,amsmath,amsthm,amsfonts}

\usepackage[T1]{fontenc}
\usepackage[tt=false, type1=true]{libertine}
\usepackage[varqu]{zi4}
\usepackage[libertine]{newtxmath}

\usepackage[margin=1in]{geometry}
\usepackage[ruled]{algorithm2e} % For algorithms
    
    \SetAlFnt{\small}
    \SetAlCapFnt{\small}
    \SetAlCapNameFnt{\small}
    \SetAlCapHSkip{0pt} 
    \IncMargin{-\parindent}

\usepackage{enumitem}
\setlist{nosep,topsep=0pt,leftmargin=*}

\usepackage{xspace,nicefrac,graphicx,tcolorbox,xifthen,thm-restate,tikz,hyperref}

\usepackage[bb=dsserif]{mathalpha}
\hypersetup{
    colorlinks,
    allcolors=myblue
}

\usepackage{xcolor}
    \definecolor{myred}{HTML}{ea4335}
    \definecolor{mygreen}{HTML}{41a756}
    \definecolor{myblue}{HTML}{4285f4}

\usepackage[capitalize]{cleveref}
\usepackage[sortcites,sorting=nyt,style=alphabetic]{biblatex}
\usepackage{color-edits}
\addauthor{et}{myred}
\addauthor{gf}{mygreen}
\addauthor{sb}{myblue}

\newtheorem{theorem}{Theorem}[section]

\newtheorem{lemma}[theorem]{Lemma}

\newtheorem{remark}[theorem]{Remark}

\theoremstyle{definition}

\newtheorem{definition}{Definition}[section]

\let\a\alpha
\let\b\beta

\let\e\varepsilon

\DeclareMathOperator*{\argmax}{\arg\max}

\newcommand{\Ex}[2][]{\mathop{\mathbb{E}}_{#1}\left[#2\right]}
\renewcommand{\Pr}[2][]{\mathop{\mathbb{P}}_{#1}\left[#2\right]}
\newcommand{\One}[1]{\mathbb{1}\left[#1\right]}
\usepackage{xifthen}
\newcommand{\Line}[4]{%
    #1 &\;#2\;#3\ifthenelse{\isempty{#4}}{}{&&\qquad\left(#4\right)}%
}

\newcommand{\type}{\theta}
\newcommand{\mechanism}{\hyperref[algo:algo]{\color{black}First-Price Pseudo-Auction with Multi-Round Reserves\xspace}}
\newcommand{\policy}{\hyperref[def:policy]{\color{black}Robust Bidding Policy}\xspace}
\newcommand{\kmax}{k_{\max}}
\newcommand{\vstar}{v^{\star}}
\newcommand{\req}{\texttt{Req}}

\newcommand{\Blk}{\texttt{Blk}}

\newcommand{\citewithauthor}[1]{\textcite{#1}}

\newenvironment{myproof}[1][Proof.]{%
    \begin{proof}[#1]%
}{%
    \end{proof}%
}

\title{Robust Pseudo-Markets for Reusable Public Resources}

\author{
    Siddhartha Banerjee
    \thanks{Supported in part by AFOSR grant FA9550-23-1-0068, ARO MURI grant W911NF-19-1-0217, NSF grants ECCS-1847393 and CNS-195599, and the Simons Institute for the Theory of Computing.}\\
    Cornell University\\
    \texttt{sbanerjee@cornell.edu}
    \and
    Giannis Fikioris
    \thanks{Supported in part by the Department of Defense (DoD) through the National Defense Science \& Engineering Graduate (NDSEG) Fellowship, the Onassis Foundation -- Scholarship ID: F ZS 068-1/2022-2023, and AFOSR grant FA9550-23-1-0068.}\\
    Cornell University\\
    \texttt{gfikioris@cs.cornell.edu}
    \and
    \'Eva Tardos
    \thanks{Supported in part by AFOSR grants FA9550-19-1-0183 and FA9550-23-1-0068.}\\
    Cornell University\\
    \texttt{eva.tardos@cornell.edu}
}

\date{\vspace{-25pt}}

\begin{document}

\maketitle

\begin{abstract}
    We study non-monetary mechanisms for the fair and efficient allocation of \emph{reusable public resources}, i.e., resources used for varying durations. We consider settings where a limited resource is repeatedly shared among a set of agents, each of whom may request to use the resource over multiple consecutive rounds, receiving utility only if they get to use the resource for the full duration of their request. Such settings are of particular significance in scientific research where large-scale instruments such as electron microscopes, particle colliders, or telescopes are shared between multiple research groups; this model also subsumes and extends existing models of repeated non-monetary allocation where resources are required for a single round only.

We study a simple pseudo-market mechanism where upfront we endow each agent with some budget of artificial credits, with the budget proportion reflecting the \emph{fair share} of the resource we want the agent to receive. The endowments thus define for each agent her \emph{ideal utility} as that which she derives from her favorite allocation with no competition, but subject to getting at most her fair share of the resource across rounds.
Next, on each round, and for each available resource item, our mechanism runs a first-price auction with a selective reserve, wherein each agent submits a desired duration and a per-round-bid, which must be at least the reserve price if requesting for multiple rounds; the bidder with the highest per-round-bid wins, and gets to use the item for the desired duration. 
We consider this problem in a Bayesian setting and show that under a carefully chosen reserve price, irrespective of how others bid, each agent has a simple strategy that guarantees she receives a $1/2$ fraction of her \textit{ideal utility} in expectation. 
We also show this result is tight, i.e., no mechanism can guarantee that all agents get more than half of their ideal utility.
\end{abstract}

\section{Introduction}

Our goal in this paper is to design mechanisms that enable fair and efficient utilization of reusable public resources between agents who share the resource. To formalize what we mean, it helps to consider the problem faced by researchers sharing some expensive scientific equipment, such as a telescope, mass spectrometer, gene sequencer, etc.
Such settings exhibit several common features: 
\begin{itemize}
    \item \emph{Resource constraints:} A telescope can at any time be used only by one researcher; a biology core facility may have multiple spectrometers allowing simultaneous sharing by a few, but still, not by all agents. This necessitates some form of centralized coordination.
    \item \emph{Stochastic and time-sensitive demands:} Research is uncertain, and so a researcher may not always know beforehand when they may need to use the equipment. Moreover, requirements are often time-sensitive, and cannot be delayed. For example, astronomers often use less powerful telescopes to find potential astronomical events of interest; as soon as one is detected they need a large-scale telescope to take detailed measurements before the event ends. Consequently, the coordinator of the larger telescope needs to make allocation decisions on the fly while being uncertain about future events. 
    \item \emph{Multi-round demands:} Different researchers may require to use the instrument for different lengths of time to complete their experiments. Moreover, they may not be able to interrupt them and/or resume later. For example, stopping a gene sequencer renders the samples and chemicals unusable. In other cases, a very large setup time makes switching between tasks not viable, as is the case in of focusing a large telescope towards a specific part of the sky. Hence, the coordinator may need to enable reserving the resource for long durations that block the workflow of some researchers. 
    \item \emph{Strategic behavior by agents:} At the end of the day, given any coordination mechanism, competing agents will try to `game' the mechanism for their own good, and there is always a danger that this may lead to a tragedy of the commons, where the resource is inefficiently utilized.
\end{itemize}

The way these challenges are handled in practice is often ad-hoc. For example, modern large-scale telescopes such as the \href{https://webb.nasa.gov/}{James Webb telescope}~\cite{jameswebb} use complex protocols for time-sharing between researchers, based on a combination of guaranteed slots, proposal-based allocations, and on-demand slots. Of course, one common solution to any such coordination problem is to enable some form of `free market', and indeed, astronomers have proposed using credit-based market mechanisms~\cite{etherton2004free,saunders2018abstract}. However, our understanding of such non-monetary mechanisms is limited, especially when incorporating dynamics, uncertainty, and scheduling constraints. Our work aims to advance our understanding of such \emph{pseudo-market} mechanisms for such settings.

At a high level, it seems clear that the principal who owns the instrument should decide when each agent gets to control the instrument in a way that aims for high overall utilization while ensuring equitable distribution of resource usage between agents. One way to do so could be via round-robin sharing -- however, while this is clearly equitable and maximizes utilization, in many cases this is undesirable as it has no alignment with when the agents would most benefit from the instrument, leading to low utility guarantees for the agents.
Unfortunately, without charging money, it is unclear how the principal can determine the exact utility that each agent gets from using the instrument on any given round. The principal could of course ask agents to report their utilities, but without money, the principal can not incentivize agents to report truthfully. Additionally, there is no direct way to compare these reports since there is no relative scale of their utilities. 

Pseudo-markets based on artificial currencies offer a way to guarantee fairness and ensure that the instrument is used in a way that generates high utilities for the agents.
The basic idea is that the principal first endows each agent with some budget of artificial credits proportional to the \emph{fair share} of the resource each agent is entitled to receive. These fair shares are exogenously specified, in a way that reflects the ex-ante `priority/importance' accorded to each agent. After the initial allocation of credits, the principal uses some mechanism to allocate the resource to some agents, charging them in this artificial currency. In the case that the allocation process happens only once, running a first-price auction offers an appealing mechanism, where the equilibrium being the outcome of the corresponding Fisher market. 
\cite{DBLP:conf/sigecom/GorokhBI21} studied the case when allocation happens repeatedly over a number of rounds but the item is required only for a single round at a time. They study a first-price mechanism and show approximate fairness properties of the resulting allocation. 
Their approach however depends on having only single-round allocations, and as we demonstrate below, performs poorly in settings that require multi-round allocations. Consequently, we need new ideas and techniques to extend their robustness notion to settings where the resources may be needed for varying amounts of time.

\subsection{Overview of our results and techniques}

Our basic setting is as follows: A principal has a single item that is to be shared between a set of agents over $T$ rounds. In each round, each agent has a random requirement which is comprised of a required duration, and a value for utilizing the item over that duration. 
The principal uses a pseudo-market to coordinate the agents, where she first endows each agent with some budget of artificial credits, and then whenever the resource is free, runs some mechanism to determine who should get to use it, and for how long. 

Our aim is to give \emph{per-agent performance guarantees} under minimal behavioral assumptions on other agents. 
In particular, following~\cite{DBLP:conf/sigecom/GorokhBI21}, we define the \textit{ideal utility} of an agent with fair share $0 < \a < 1$ as the best long-run average utility she can achieve in a setting without competition, but where she is constrained to have the item for at most an $\a$ fraction of the rounds (see \cref{sec:ideal} for the formal definition). The ideal utility is defined independently of other agents' demands or behavior, thus making it a reliable measure of the maximum utility an agent can attain with their fair share.
Our results focus on what fraction of her ideal utility an agent can achieve in a \emph{minimax} sense, i.e., irrespective of how other agents bid. 
This is in contrast to other measures like no-regret, where the resulting utility is compared with a benchmark that depends on how other agents bid.
In this sense, our guarantees can be viewed as characterizing the robustness of the underlying mechanism with respect to an agent's ideal utility.

To better appreciate our mechanism's guarantees, it is instructive to first observe that simple mechanisms like round-robin have poor performance, even when agents have single-round demands. Consider a setting with $n$ agents, each having fair share $\a = \nicefrac 1 n$ and non-zero value with probability $\nicefrac 1 n$. Every round, the probability that round-robin allocates the item to an agent with positive value that round is only $\nicefrac 1 n$, even though the probability of some agent having positive value is $\Omega(1)$. Thus, every agent receives a $O(1/n)$ fraction of her ideal utility in expectation, even though our mechanism guarantees a $\Omega(1)$ fraction.

In this regard, our main contribution is \textbf{\mechanism}, a new mechanism that guarantees that every agent can realize at least $\nicefrac{1}{2}$ of her ideal utility, even with multi-round allocations.
In the setting of single-round allocations, this same guarantee is achieved by a simple first price auction~\cite{DBLP:conf/sigecom/GorokhBI21}; however, this auction performs poorly with multi-round reservations, even if agents are truthful.
The reason for this is that if we allow agents to reserve the item for longer rounds, in a round in which there is no serious demand, an agent can capture the item for a long time at a very low price, and such a reservation can block higher-valued demands that arrive in later rounds. 

To overcome this difficulty, our mechanism uses a reserve price for multi-round reservations.
In each round each agent can request the resource for a consecutive sequence of rounds starting with the current one by submitting a requested duration and their per-round bid of that duration. Bids that request the resource for more than one round need to be at least the reserve price; aside from that rule, the agent with the highest per-round bid wins.
We can think of our mechanism as a combination of a spot market for the current round, and a buy-ahead reservation option but with a price floor. 

Our main technical contributions are as follows:
\begin{itemize}
    \item In \cref{sec:guar} we study our mechanism, where an agent with fair share $\a$ is endowed with budget $\a T$ and the other agents have a total budget of $(1-\a)T$.
    Now when the reserve price is $r$, we prove that an agent with ideal utility $\vstar$ can guarantee $\vstar T \min\{ 1/r, 1-1/r \} - O(\sqrt T)$ utility in expectation over $T$ rounds, regardless of how the other agents behave (\cref{thm:guar:guarantee}). This quantity is maximized when $r = 2$, in which case the agent can guarantee half her ideal utility. We then show that this is the best possible bound in our mechanism: an agent cannot guarantee more than $\vstar T (\min\{ 1/r, 1-1/r \})^+$ expected utility (\cref{thm:guar:impossibility}). This shows that reserve prices are essential in our mechanism for multi-round reservations.

    The $1/r$ part in the minimum of both results comes from the fact that an agent with budget $\a T$ and multi-round demands cannot get the item for more than $\a T / r$ rounds when the reserve price is $r$. The argument for being able to guarantee the $1-1/r$ part is the following: The other agents have budget at most $T$ which means they can win at most $T/r$ rounds with bids at or above the reserve price, leaving $T(1 - 1/r)$ for the agent if she is willing to pay the reserve price. However, with the multi-round demands, the rounds are not at all independent. For example, if the adversary could reserve every other round ahead of time, this could eliminate all value for an agent with only two-round demands. Using a martingale argument we prove that because the other agents' behavior is independent of the value of the agent, the agent can get $\vstar$ utility from each one of those rounds in expectation. 
    
    For the upper bound result, the other agents can get the item for $T/r$ rounds, and this leaves the agent with $T - T/r$ rounds in which the item is available. We prove that if the mechanism allows long reservation durations, then our agent may have high value in only an $\a$ fraction of the available rounds.

    \item In \cref{sec:hardness} we prove that no non-monetary mechanism (pseudo-market or otherwise) can guarantee that every agent can get more than half her ideal utility (\cref{thm:imposs}), making our previous result optimal. We prove this by examining an example where every agent has positive demand with low probability, that lasts many rounds.

    \item In \cref{sec:multi} we study the same setting we did in \cref{sec:guar}, but when there are $L$ identical items that the principal can allocate to the agents (we still assume each agent wants at most one item at a time). In this setting the ideal utility of an agent with fair share $\a$ allows her to have the item for a fraction of $\a L$ rounds. We show that the agent can again guarantee half of her ideal utility (\cref{thm:mult:guarantee}).
\end{itemize}

\subsection{Paper Outline}

Before presenting our work, we survey related literature in~\cref{sec:litrev}.
In \cref{sec:setting}, we define the simplest case of the reusable resource model, where the principal has a single resource to allocate in each round, and outline the general structure of the pseudo-market mechanism that we study in this work. For most of the paper, we focus on this single resource setting: we first define our per-agent benchmarks (\cref{sec:ideal}) and then state our main robustness guarantees (\cref{sec:guar}) and associated hardness results (\cref{sec:hardness}). We extend our basic setting to incorporate multi-unit settings in \cref{sec:multi}. 

\section{Related Literature}
\label{sec:litrev}

Our work sits at the intersection of two topics in mechanism design: (i) non-monetary mechanisms for resource allocation, and (ii) dynamic mechanisms with state.
%, and (iii) non-equilibrium welfare guarantees. 
We now briefly summarize the literature on these topics.

Classical non-monetary mechanism design utilizes a wide variety of models and objectives to realize good welfare outcomes despite having strategic agents. Some of these include targeting alternate solution concepts such as pairwise stability~\cite{roth1992}, disregarding incentives to focus on fairness properties of realized outcomes~\cite{moulin2002proportional,caragiannis2016unreasonable,banerjee2022online}, partial public information (e.g., utilities are public but feasibility is private~\cite{arpitaShaddin}), designing lotteries to approximate efficient outcomes~\cite{moulin2007scheduling,procaccia2013approximate,kojima2010incentives}, explicitly hurting efficiency to align incentives (i.e., `money burning'~\cite{hartline2008optimal,DBLP:conf/sigecom/ColeGG13}), and using ex-post verification~\cite{branzei2015verifiably}. Most of this literature, however, deals with one-shot (i.e., static) allocation settings. 
%\etcomment{added a reference, can we force latex to put them in the other order?}\sbcomment{Done!}

More recently, there has been an increasing focus on \emph{pseudo-markets} -- simulating monetary mechanisms using an \emph{artificial currency} -- driven largely by their success in real-world deployment for university course allocation~\cite{budish2017course}, food banks~\cite{walsh2014allocation,prendergast2022allocation} and cloud computing platforms~\cite{dawson2013reserving,vasudevan2016customizable}.
Theoretical foundations of such mechanisms have been studied in the context of one-shot combinatorial assignment problems~\cite{budish2011combinatorial,2017nashwelfare}, but also in dynamic settings, including redistribution mechanisms~\cite{cavallo2014incentive}, and approximate mechanisms for infinite-horizon Bayesian settings with knowledge of value distributions~\cite{jackson,guo2010,balseiro2017,gorokh2017}. The latter works all build on the core idea of \citewithauthor{jackson} of `linking' multiple allocation problems to mitigate any gains from strategic behavior in any one problem. More recently,~\citewithauthor{DBLP:conf/sigecom/GorokhBI21} showed how pseudo-markets can be used for repeated single-round allocations to get individual performance guarantees \emph{without} knowing demand distributions; our work adapts and extends their ideas to the more complex reusable resource setting.

The challenge of dealing with both fixed budgets and reusable resources also places our work in the active area of dynamic mechanisms with state; once the resource is allocated, it is unavailable for the next few rounds, and the user with the allocation has decreased credit. 
The difficulty in these problems arises due to factors that couple allocations across time; for example, incomplete
information and learning~\cite{DBLP:conf/wine/KanoriaN14,Iyer2014,devanur2015perfect,nekipelov2015econometrics},
cross-round constraints including budget limits~\cite{nazerzadeh2008,gummadi2012repeated,leme2012sequential,balseiro2015repeated}, 
leftover packets in a queuing system \cite{DBLP:conf/sigecom/GaitondeT20,DBLP:conf/sigecom/GaitondeT21}, 
stochastic fluctuations in the underlying
setting~\cite{gershkov2009dynamic,bergemann2010dynamic,bergemann2011dynamic}, 
adversarial environments \cite{DBLP:journals/mansci/BalseiroG19}, etc. 
Analyzing equilibria in repeated settings however can be difficult, and so authors have explored approximation techniques such as mean-field approaches~\cite{gummadi2012repeated,balseiro2015repeated} and bi-criteria approximations~\cite{nekipelov2015econometrics}.

Another relevant and recent work is that of \cite{gaitonde2022budget}, where value maximizing agents have budget constraints that correspond to real money. They show a regret type guarantee against the strategy that spends at most the agent's average budget in expectation each round, but their result degrades a lot when the behavior of the other agents is adversarial. \cite{balseiro2021landscape,deng2021towards,pai2014optimal} study a similar setting where the agents' behavior is assumed to reach equilibrium; the first and third focusing on revenue maximization and the second on welfare maximization.

%\gfedit{
%The idea of linking decisions can also be used for allocating in multiple rounds over time. This idea is explored for single-item allocation and symmetric agents in a line of work starting with~\cite{guo2009}, and refined by~\cite{balseiro2017} (this idea has also been extended to general allocation settings~\cite{gorokh2017}; more on this below). These works assume an endogenously provided way of combining preferences of agents into a welfare function and strive to maximize welfare subject to incentive compatibility. There are several major differences that set us apart from this work: 1) in line with literature in economics, we avoid agglomerating utilities of agents into a welfare function in the absence of transfers, as it can be challenging to rigorously justify doing so; 2) the allocation setting we consider here is considerably more general; 3) our guarantees hold for the maxmin value, rather than using the Nash equilibrium as a solution concept; and 4) we concentrate on the performance of simple, practically implementable mechanisms.

%Another closely related line of work explores the properties of artificial currency (i.e., \emph{scrip}) economies~\cite{kash2007optimizing,halpern2008beyond}.
%}
\section{Reusable Public Resources and Pseudo-Markets -- Basic Setting}
\label{sec:setting}

We now formally define the simplest case of the reusable resource model, where the principal has a single resource it can allocate in each round -- for example, time-sharing a single telescope. We then outline the general structure of pseudo-market mechanisms that we study in this work. In \cref{sec:ideal,sec:guar,sec:hardness}, we focus on this setting; subsequently, we extend our model to incorporate multi-unit settings in \cref{sec:multi}.

\subsection{Allocating a Single Reusable Public Resource}

There are $n$ agents and $T$ rounds.
In each round, the principal has a single item to allocate.
Agents have single-minded multi-round valuations; formally, in every round $t\in [T]$, each agent $i\in [n]$ samples a random \emph{type} $\type_{i}[t]=(V_{i}[t], K_{i}[t])$, where $K_{i}[t]$ is the number of rounds that agent $i$ needs the item for starting from round $t$, and $V_{i}[t]$ is the \emph{per-round value} she gets if she is allocated the item for the next $K_{i}[t]$ rounds. 
In other words, if agent $i$ is allocated the item on rounds $t, t+1, \ldots, t+K_{i}[t]-1$ (henceforth denoted $[t,t+K_{i}[t]-1]$), then she receives a total utility of $K_{i}[t]V_{i}[t]$; on the other hand, if the agent is not allocated the item for all of these rounds, then she does not get any utility arising from her round $t$ demand.
We henceforth use $\Theta=\mathbb{R}_+\times\mathbb{N}$ to denote the type space for each agent and round.

We assume that each agent $i$ in each round $t$ draws demand type $\type_{i}[t]$ from some underlying distribution $\mathcal F_i$, that is \emph{independent across agents} and \emph{across rounds}.
In particular, agent $i$'s demand $\type_{i}[t]$ is drawn independently in round $t$ irrespective of her demand in previous rounds.
Note that this means if agent $i$ has $K_{i}[t] > 1$ but is not allocated the item in round $t$, then in round $t+1$, the earlier demand is lost, and she draws a new demand $\type_{i}[t+1]$.
Having a demand that is lost if not immediately allocated is meaningful in many settings where it is not possible for agents to hold back on executing a demand till a later round.
For example, a biologist may not be able to preserve a sample that she needs a microscope to study. Similarly, an astronomer might need to immediately redirect a shared telescope to point toward a transient phenomenon she wants to observe. In addition, even when the astronomer is denied use of the telescope to take measurements of any particular event, new opportunities may arise soon afterward.

Once an agent $i$ is allocated the item for days $[t, t+K_{i}[t]-1]$, we assume the item is unavailable for reallocation to \emph{any} agent (including agent $i$). In other words, in every round $t$, either the item is unavailable for allocation, or the principal commits it to some agent $i$ for rounds $[t,t+K_{i}[t]-1]$. 
The commitment to not interrupt allocation models situations with large setup cost. For example, there is often some setup cost in preparing an instrument for an experiment, or in directing and focusing a telescope, and so it may be desirable for agents to complete any task they start.

Finally, we assume that each agent $i$ has a fair share $\a_i$, where $\sum_i \a_i = 1$. Fair shares are usually exogenously defined and represent the fraction of the resource that the principal wants each agent to have (which then determines her associated \emph{ideal utility}, see \cref{sec:ideal}).

\subsection{Artificial Currency Mechanism for Reusable Resources}
\label{ssec:mechanism}

Given the above setting, the mechanism we study for allocating the resource is a \emph{pseudo-market} (or artificial credit) mechanism. The basic idea behind such mechanisms is to endow all agents competing for the shared resource with some budget of artificial credits, which they then use over time to compete in some form of repeated auction. Such mechanisms have been widely used in practice~\cite{prendergast2017food,budish2017course}, and also studied theoretically in one-shot allocation settings under known value distributions~\cite{guo2009,gorokh2017,balseiro2017}. Our presentation here is closest to that of~\cite{DBLP:conf/sigecom/GorokhBI21}, who consider the robustness properties of such mechanisms for repeated allocation with single-round demands.

Our mechanism, \mechanism, starts by endowing every agent with some artificial credits, proportional to their fair shares.
Since the currency has no intrinsic value (and hence no fixed scale), we henceforth normalize the total budget of all agents to be $T$, of which every agent $i$ has a share of $\a_i$; in other words, each agent $i$ has an initial budget $B_i[1]=\a_i T$. 
At a high level, the budget fraction $\a_i$ corresponds to the idea that agent $i$ could get to use the item on an $\a_i$ fraction of rounds.

Following the initial endowment, the basic idea behind pseudo-market mechanisms is to then run some particular mechanism in each round, and allow agents to bid (and pay) in these mechanisms using their credits. Agents have no intrinsic value for these credits (i.e., their utility is not quasi-linear, it comes only from the allocations gained), but they are unable to bid more than their remaining budget. While different works consider different mechanisms, the most commonly studied is a first-price auction~\cite{DBLP:conf/sigecom/GorokhBI21,balseiro2017,gorokh2017,guo2009}. 

Our mechanism handles multi-round allocations as follows: first, the principal declares a \emph{reserve price} $r$ for multi-round allocations; next, at the start of any round in which the resource is available, each agent declares a duration of rounds she wants to reserve the resource for, as well as a \emph{per-round} bid (which must exceed the reserve if the requested duration lasts multiple rounds). The agent with the \emph{highest per-round bid} is then awarded the item for her requested duration and is `charged' her bid times the duration from her credit budget.

We note that if we set a reserve price $r > 1$, then, if only multi-round requests are made, there are not enough credits among the agents to enable the resource to be allocated in every round. 
However, as we show in \cref{thm:guar:impossibility}, this wastage is necessary with multi-round demands to obtain any meaningful performance guarantee. Note though that in settings where preemption is not possible, even if the resource is used by a single agent with full knowledge of all future demands, then also the agent might leave the resource unused at times, so as not to interfere with future higher-valued demand.

We present the mechanism in detail in \cref{algo:algo}. In~\cref{sec:multi}, we discuss how to extend it to settings where the principal has more than one resource to allocate.

\begin{algorithm}[t]
\SetAlgoNoLine
\caption{\mechanism}
\label{algo:algo}
\KwIn{Rounds $T$, agents $n$, reserve $r \ge 0$, and agents' fair shares $\{\a_i\}_{i\in[n]}$} 
%(i.e., $B_i[1] = \a_i T\,\forall\,i\in [n]$)}
\textbf{Initialize} $t = 1$, agent budgets $B_i[1] = \a_i T\quad\forall\,i\in [n]$\;
\While{$t \le T$}
{
    Collect bids $b_{1}[t], \ldots, b_{n}[t]$ and desired durations $d_{1}[t], \ldots, d_{n}[t]$\; 
    Let $\mathcal V = \big\{ i\in [n] : b_i[t]d_i[t] \le B_i[t] \textrm{ and } (d_{i}[t] = 1 \textrm{ or } b_i[t] \ge r) \big\}$%
    \tcp*{Determine valid bids}

    \eIf{$\mathcal V=\emptyset$}{
        Do not allocate item and set $t=t+1$\;
    }{
        Define $I_t = \argmax_{i\in\mathcal V} b_{i}[t]$ (ties broken arbitrarily)%
        \tcp*{Choose winning agent}
    
        Update $B_{i}[t+1] = B_{i}[t] - b_{i}[t]d_{i}[t]\One{i=I_t}$%
        \tcp*{Update agents' budgets}
        
        Allocate item to agent $I_t$ for rounds $[t, t + d_{I_t }[t] - 1]$\;
        
        Set $t = t + d_{I_t }[t]$%
        \tcp*{Block item for requested duration}
    }
}
\end{algorithm}

\section{Individual Agent Benchmarks: Fair Shares and the Ideal Utility}
\label{sec:ideal}

In this section, we define the utility benchmarks we consider for the agents. When mechanisms can use real money and agents have quasi-linear utilities, then payments provide an easy way to compare different agents' values and utilities. In contrast, when there are no payments that affect the agents' utilities, then there is no way to make interpersonal comparisons between agents. 
For this reason, we need a welfare benchmark for each agent that is independent of other agents' values. To this end, we adapt an idea from~\cite{DBLP:conf/sigecom/GorokhBI21} (which in turn borrows ideas from the bargaining literature and the Fisher market model), wherein agents' benchmarks are defined by their (exogenous) fair shares as well as their own relative valuations for items in different rounds.

The main idea behind our benchmark is that for each agent $i$, her budget fraction $\a_i$ (with $\sum_j \a_j = 1$) determines the \textit{fair share} of the overall resource, i.e., the fraction of total rounds she is entitled to utilize while respecting the rights of other agents to access the resource. 
To see how this translates into our welfare benchmark, consider the following simple example: suppose we have $n$ agents, where each agent has fair share $\a_i = 1/n$. Moreover, suppose every agent $i$ has $(V_{i}[t], K_{i}[t]) = (1, 1)$ with probability $1$ in every round. An agent's maximum total utility \emph{without competition} is $T$, but it would be unreasonable to expect this to be attainable for any agent. In contrast, by symmetry, each agent could expect to win $T/n$ total rounds resulting in $T/n$ total utility, which is indeed easily achieved (for example, via a round-robin allocation scheme).

\citewithauthor{DBLP:conf/sigecom/GorokhBI21} extend the above idea to define the \textit{ideal utility} for agents in settings where every request lasts for just one round. Their basic definition asserts that an agent's ideal utility is \emph{the highest per-round utility she can get while ensuring that other agents can get at least their fair share of the resource}.
Formally, for each agent $i$, they consider a simplified setting with no other agents, but where agent $i$ is constrained to request the item for at most an $\a_i$ fraction of the rounds, and define agent $i$'s ideal utility to be the maximum expected per-round utility she can achieve in this setting. 
For example, if agent $i$ has $(V_{i}[t], K_{i}[t]) = (1, 1)$ with probability $1$, then her ideal utility is thus $\a_i$ for any fair share $\a_i$; more generally, if the agent has value $V_{i}[t]\sim\mathcal{F}_i$ (and $K_i[t]=1$) and fair share $\a_i$, her ideal utility essentially corresponds to that achieved by requesting the item only on rounds in which the agent's value $V_i[t]$ is in the top $\a_i$ quantile of her value distribution, which makes the agent request the item with probability $\a_i$ depending on her demand. 

A first challenge in extending the ideal utility to our setting with reusable resources is that now it is tricky to define what it means for an agent to request each round in the no-competition setting while ensuring that she is only using her fair share, since each bid may need to reserve the resource for multiple days.
To this end, given fair share $\a_i$, we define agent $i$'s ideal utility to be her \emph{long-run average utility in an infinite horizon setting with no competition, subject to her long-run average resource utilization being at most $\a_i$}.

In more detail, let $\pi:\Theta\to[0,1]$ denote a (stationary) policy that specifies for each type $\theta = (V, K)$ the probability with which the agent requests to reserve the resource for $K$ rounds, conditioned on it being available. The agent's ideal utility (rate) is that which she obtains under the optimal policy $\pi$ (which is uniquely defined under mild technical conditions). Note though that requesting the item in any round affects her ability to request it in future rounds, even with no competition. To formalize this, let $Z[t] = 1$ denote that the agent chooses to reserve the item in round $t$ for $K[t]$ rounds; $Z[t] = 0$ otherwise.
This affects the future availability of the item: if $Z[t] = 1$ then the item is unavailable for the next $K[t] - 1$ rounds. 
Let $A[t]$ be an indicator variable for item availability, with $A[t]=1$ if the item is available in round $t$, and $A[t] = 0$ if unavailable; then, if $Z[t] = 1$, we have that $A[t'] = 0$ for all $t'\in[t+1, t+K[t]-1]$.
Now, given policy $\pi$, we can define $Z[t]$ as a Bernoulli random variable that is $1$ with probability $A[t]\pi(\type[t])$.
Note though that these definitions are not recursive: even though the definition of $Z$ involves $A$ and vice versa, $A[t]$ depends only on $Z[t']$ for $t' < t$ and $Z[t]$ depends only on $A[t]$.
%\gfdelete{Formally, let $A[t]\in\{0,1\}$ denote the state of the resource at time $t$ (where $A[t]=1$ indicates the resource is available), and $\pi:\Theta\to[0,1]$ denote a (stationary) policy that specifies for each type $\theta=(V, K)$ the probability with which the agent requests to reserve the resource for $K$ rounds conditioned on it being available.}

Now we can make the following definition:
 
\begin{definition}[Ideal Utility]\label{def:ideal:single}
    Consider the single reusable resource setting, with a single agent $i$ with fair share $\a_i$ and type $\type[t] = (V[t], K[t])$ in round $t$ drawn independently of other rounds from distribution $\mathcal F_i$.
    For any policy $\pi:\Theta\rightarrow [0,1]$, let $Z[t]\sim \text{Bernoulli}(A[t]\pi(\theta[t]))$ denote a sequence of indicator variables each of which is $1$ in round $t$ if the resource is available (indicated by $A[t] = 1$) and is requested by the agent, else $0$; moreover if $Z[t]=1$, then $A[t'] = 0$ for all $t'\in[t+1, t+K[t]-1]$. 
    Now, the \textit{ideal utility $\vstar_i$} for agent $i$ is defined as the value of the following constrained infinite-horizon control problem:
    \begin{equation}
    \label{eq:ideal:MDP}
    \begin{split}
        \max{}_{\pi} \quad &\lim_{H\rightarrow\infty}\frac{1}{H}\sum_{t=1}^H V[t]K[t]Z[t]
        \\
        \textrm{such that} \quad
         &\lim_{H\rightarrow\infty}\frac{1}{H}\sum_{t=1}^H K[t]Z[t] \le \a_i
    \end{split}
    \end{equation}
\end{definition}

Note that the above problem does not depend on our true (finite) horizon $T$. Moreover, assuming $V[t], K[t]$ are bounded, via the Markov chain ergodic theorem we have that for any policy $\pi$, the above time average costs exist and equal their expected value under the stationary distribution of the resulting Markov chain. However, unlike a standard average cost MDP, due to the additional constraint, the optimal policy here may not be deterministic.

Intuitively, the above definition extends the notion of the ideal utility to the single reusable resource setting by again considering a world with only a single agent $i$ with fair share $\a_i$, and allowing the agent to choose any stationary request policy (i.e., the probability, as a function of $\type[t]$, with which the agent can reserve the resource whenever it is free) subject to it using the resource at most an $\a_i$ fraction of rounds on average.
Defining the control problem over the infinite horizon allows us to ignore boundary issues (e.g., if there are $5$ rounds remaining but agent $i$'s demand lasts $6$ rounds); moreover, note that in the case of single-round demands, our definition recovers that of~\citewithauthor{DBLP:conf/sigecom/GorokhBI21}.

\subsection{Computing the ideal utility}
\label{ssec:comp}

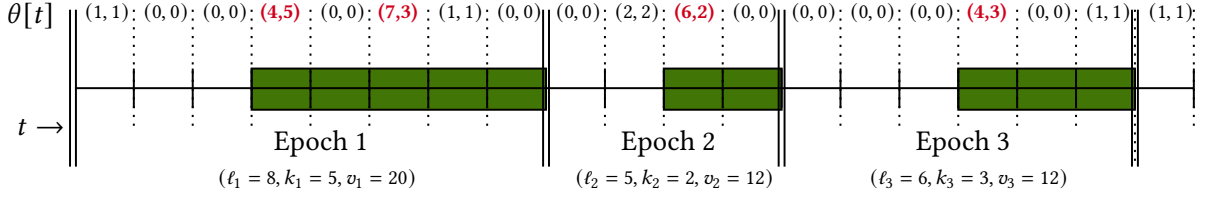
\begin{figure}[t]
    \centering
    \scalebox{0.99}{\tikzset{every picture/.style={line width=0.75pt}} %set default line width to 0.75pt        

\begin{tikzpicture}[x=0.75pt,y=0.75pt,yscale=-1,xscale=1]

%Shape: Rectangle [id:dp8722661595401164] 
\draw  [fill={rgb, 255:red, 65; green, 117; blue, 5 }  ,fill opacity=1 ] (510,150) -- (600,150) -- (600,171) -- (510,171) -- cycle ;
%Shape: Rectangle [id:dp9984712508772009] 
\draw  [fill={rgb, 255:red, 65; green, 117; blue, 5 }  ,fill opacity=1 ] (360,150) -- (420,150) -- (420,171) -- (360,171) -- cycle ;
%Shape: Rectangle [id:dp9919236109151666] 
\draw  [fill={rgb, 255:red, 65; green, 117; blue, 5 }  ,fill opacity=1 ] (150,150) -- (300,150) -- (300,171) -- (150,171) -- cycle ;
%Straight Lines [id:da4752460053363434] 
\draw    (60,160) -- (630,160) ;
%Straight Lines [id:da7340069332403847] 
\draw    (90,150) -- (90,170) ;
%Straight Lines [id:da08907364329824019] 
\draw    (120,150) -- (120,170) ;
%Straight Lines [id:da6228671748667489] 
\draw    (180,150) -- (180,170) ;
%Straight Lines [id:da11304241008547522] 
\draw    (210,150) -- (210,170) ;
%Straight Lines [id:da686015268836075] 
\draw    (150,150) -- (150,170) ;
%Straight Lines [id:da5948160280639214] 
\draw    (270,150) -- (270,170) ;
%Straight Lines [id:da3777815653483254] 
\draw    (300,150) -- (300,170) ;
%Straight Lines [id:da7930574376741437] 
\draw    (240,150) -- (240,170) ;
%Straight Lines [id:da782831926705017] 
\draw    (360,150) -- (360,170) ;
%Straight Lines [id:da6486640746820416] 
\draw    (390,150) -- (390,170) ;
%Straight Lines [id:da9771857952029788] 
\draw    (330,150) -- (330,170) ;
%Straight Lines [id:da5308445129073329] 
\draw    (450,150) -- (450,170) ;
%Straight Lines [id:da085975368364265] 
\draw    (480,150) -- (480,170) ;
%Straight Lines [id:da7374730320522569] 
\draw    (420,150) -- (420,170) ;
%Straight Lines [id:da25005601848998227] 
\draw    (540,150) -- (540,170) ;
%Straight Lines [id:da480965439187123] 
\draw    (570,150) -- (570,170) ;
%Straight Lines [id:da3239537970177473] 
\draw    (510,150) -- (510,170) ;
%Straight Lines [id:da7174081042324254] 
\draw    (600,150) -- (600,170) ;
%Straight Lines [id:da18274171868892752] 
\draw    (630,150) -- (630,170) ;
%Straight Lines [id:da36728723501090754] 
\draw    (301.5,120) -- (301.5,200)(298.5,120) -- (298.5,200) ;
%Straight Lines [id:da609903223858034] 
\draw  [dash pattern={on 0.84pt off 2.51pt}]  (600,120) -- (600,200) ;
%Straight Lines [id:da3113399464457707] 
\draw  [dash pattern={on 0.84pt off 2.51pt}]  (90,120) -- (90,180) ;
%Straight Lines [id:da28471099667352884] 
\draw  [dash pattern={on 0.84pt off 2.51pt}]  (120,120) -- (120,180) ;
%Straight Lines [id:da5043060855902493] 
\draw  [dash pattern={on 0.84pt off 2.51pt}]  (150,120) -- (150,180) ;
%Straight Lines [id:da7957409080989553] 
\draw  [dash pattern={on 0.84pt off 2.51pt}]  (180,120) -- (180,180) ;
%Straight Lines [id:da27804776058783776] 
\draw  [dash pattern={on 0.84pt off 2.51pt}]  (210,120) -- (210,180) ;
%Straight Lines [id:da27964971638397573] 
\draw  [dash pattern={on 0.84pt off 2.51pt}]  (240,120) -- (240,180) ;
%Straight Lines [id:da18505086745110289] 
\draw  [dash pattern={on 0.84pt off 2.51pt}]  (270,120) -- (270,180) ;
%Straight Lines [id:da7983062104747054] 
\draw  [dash pattern={on 0.84pt off 2.51pt}]  (330,120) -- (330,180) ;
%Straight Lines [id:da13179403506957033] 
\draw    (601.5,120) -- (601.5,200)(598.5,120) -- (598.5,200) ;
%Straight Lines [id:da2270079737227393] 
\draw  [dash pattern={on 0.84pt off 2.51pt}]  (360,120) -- (360,180) ;
%Straight Lines [id:da4538938326625028] 
\draw  [dash pattern={on 0.84pt off 2.51pt}]  (390,120) -- (390,180) ;
%Straight Lines [id:da4625493550099009] 
\draw    (421.5,120) -- (421.5,200)(418.5,120) -- (418.5,200) ;
%Straight Lines [id:da240876823637989] 
\draw  [dash pattern={on 0.84pt off 2.51pt}]  (450,120) -- (450,180) ;
%Straight Lines [id:da8771349989264556] 
\draw  [dash pattern={on 0.84pt off 2.51pt}]  (480,120) -- (480,180) ;
%Straight Lines [id:da19138621439776915] 
\draw  [dash pattern={on 0.84pt off 2.51pt}]  (510,120) -- (510,180) ;
%Straight Lines [id:da4623919307095856] 
\draw  [dash pattern={on 0.84pt off 2.51pt}]  (540,120) -- (540,180) ;
%Straight Lines [id:da23292237380713576] 
\draw  [dash pattern={on 0.84pt off 2.51pt}]  (570,120) -- (570,180) ;
%Straight Lines [id:da6490192319500108] 
\draw  [dash pattern={on 0.84pt off 2.51pt}]  (630,120) -- (630,180) ;
%Straight Lines [id:da08739798629519013] 
\draw    (60.5,120) -- (60.5,200)(57.5,120) -- (57.5,200) ;

% Text Node
\draw (63,115) node [anchor=north west][inner sep=0.75pt]   [align=left] {{\scriptsize $(1,1)$}};
% Text Node
\draw (93,115) node [anchor=north west][inner sep=0.75pt]   [align=left] {{\scriptsize $(0,0)$}};
% Text Node
\draw (123,115) node [anchor=north west][inner sep=0.75pt]   [align=left] {{\scriptsize $(0,0)$}};
% Text Node
\draw (153,115) node [anchor=north west][inner sep=0.75pt]  [color={rgb, 255:red, 208; green, 2; blue, 27 }  ,opacity=1 ] [align=left] {{\scriptsize \textbf{(4,5)}}};
% Text Node
\draw (183,115) node [anchor=north west][inner sep=0.75pt]   [align=left] {{\scriptsize $(0,0)$}};
% Text Node
\draw (213,115) node [anchor=north west][inner sep=0.75pt]  [color={rgb, 255:red, 208; green, 2; blue, 27 }  ,opacity=1 ] [align=left] {{\scriptsize \textbf{(7,3)}}};
% Text Node
\draw (243,115) node [anchor=north west][inner sep=0.75pt]   [align=left] {{\scriptsize $(1,1)$}};
% Text Node
\draw (273,115) node [anchor=north west][inner sep=0.75pt]   [align=left] {{\scriptsize $(0,0)$}};
% Text Node
\draw (303,115) node [anchor=north west][inner sep=0.75pt]   [align=left] {{\scriptsize $(0,0)$}};
% Text Node
\draw (333,115) node [anchor=north west][inner sep=0.75pt]   [align=left] {{\scriptsize $(2,2)$}};
% Text Node
\draw (364,115) node [anchor=north west][inner sep=0.75pt]  [color={rgb, 255:red, 208; green, 2; blue, 27 }  ,opacity=1 ] [align=left] {{\scriptsize \textbf{(6,2)}}};
% Text Node
\draw (393,115) node [anchor=north west][inner sep=0.75pt]   [align=left] {{\scriptsize $(0,0)$}};
% Text Node
\draw (423,115) node [anchor=north west][inner sep=0.75pt]   [align=left] {{\scriptsize $(0,0)$}};
% Text Node
\draw (453,115) node [anchor=north west][inner sep=0.75pt]   [align=left] {{\scriptsize $(0,0)$}};
% Text Node
\draw (483,115) node [anchor=north west][inner sep=0.75pt]   [align=left] {{\scriptsize $(0,0)$}};
% Text Node
\draw (513,115) node [anchor=north west][inner sep=0.75pt]  [color={rgb, 255:red, 208; green, 2; blue, 27 }  ,opacity=1 ] [align=left] {{\scriptsize \textbf{(4,3)}}};
% Text Node
\draw (543,115) node [anchor=north west][inner sep=0.75pt]   [align=left] {{\scriptsize $(0,0)$}};
% Text Node
\draw (573,115) node [anchor=north west][inner sep=0.75pt]   [align=left] {{\scriptsize $(1,1)$}};
% Text Node
\draw (605,115) node [anchor=north west][inner sep=0.75pt]   [align=left] {{\scriptsize $(1,1)$}};
% Text Node
\draw (24,115) node [anchor=north west][inner sep=0.75pt]    {$\type[t]$};
% Text Node
\draw (29,174.4) node [anchor=north west][inner sep=0.75pt]    {$t\rightarrow $};
% Text Node
\draw (160,180) node [anchor=north west][inner sep=0.75pt]   [align=left] {Epoch $1$};
% Text Node
\draw (337,180) node [anchor=north west][inner sep=0.75pt]   [align=left] {Epoch $2$};
% Text Node
\draw (487,180) node [anchor=north west][inner sep=0.75pt]   [align=left] {Epoch $3$};
% Text Node
\draw (131,200) node [anchor=north west][inner sep=0.75pt]  [font=\scriptsize]  {$( \ell_{1}=8,k_{1}=5,v_{1}=20)$};
% Text Node
\draw (313,200) node [anchor=north west][inner sep=0.75pt]  [font=\scriptsize]  {$( \ell_{2}=5,k_{2}=2,v_{2}=12)$};
% Text Node
\draw (463,200) node [anchor=north west][inner sep=0.75pt]  [font=\scriptsize]  {$( \ell_{3}=6,k_{3}=3,v_{3}=12)$};

\end{tikzpicture}}
    \caption{\it \small Example of an agent's ideal utility: The numbers on top are the agent's type $\type[t] = (V[t], K[t])$ each round $t$; a type in red denotes that $\req(\type[t]) = 1$. Each epoch is associated with a request that the agent made while the item was available: it includes the rounds when the agent held the item because of that request (green blocks) and the rounds before that while the item was free. For each epoch $j$ we also include its length, $\ell_j$, the number of rounds the agent holds the item for, $k_j$, and the total value the agent gets, $v_j$.}
    \label{fig:epoch}
\end{figure}

One problem with defining the ideal utility via the infinite horizon control problem in \cref{eq:ideal:MDP} is that it is unclear if it can be solved efficiently, and moreover, how to interpret the solution for any given distribution $\mathcal{F}_i$ and fair share $\a_i$. We now show how the above definition of the ideal utility for agent $i$ can be re-formulated as a simpler optimization problem, which we show can be efficiently solved by converting it into a linear program. For ease of notation, we drop the subscript $i$ for the remainder of this section.

To reformulate the above program, we first define $\req(\type)$ to be an indicator random variable that is $1$ if the agent has type $\theta$, and wants to request the item if available (in our earlier notation, for given randomized policy $\pi$, we have $\req(\type[t]))\sim\text{Bernoulli}(\pi(\theta[t]))$).
Now, given some function $\req$, we divide the entire horizon into a collection of discrete renewal cycles or \emph{epochs}, where each epoch comprises of all rounds between successive times in which the resource is released by the agent: formally, if in round $t$ the agent requests the item, then the epoch associated with that request comprises of all the rounds before $t$ since the last time the resource was unavailable (which can be $0$) and all the rounds after $t$ till the agent releases her hold of the item (i.e., rounds $[t,t+K[t]-1]$). We show an example in \cref{fig:epoch}.
Under any stationary policy $\req$, any two epochs are independent and identically distributed. Now let $q = \Pr{ \req(V, K) = 1 }$ denote the probability that the agent requests the item if it is available (where the probability is over both $(V, K) \sim \mathcal F$ and any randomization in the agent's request policy). Then the following are true for each epoch:
\begin{itemize}
    \item The number of rounds in an epoch after its start and until (and including) the round in which the agent requests for the item is distributed as $\text{Geometric}(q)$.
    
    \item If $\ell_j$ is the length of an epoch $j$, then $\Ex{\ell_j} = \nicefrac{1}{q} - 1 + \Ex{K | \req(V, K) = 1}$.
    
    \item If $v_j$ is the total utility the agent gets in epoch $j$, then $\Ex{v_j} = \Ex{V K | \req(V, K) = 1}$. Similarly, if $k_j$ is the number of rounds the agent holds the item in epoch $j$, then it holds that $\Ex{k_j} = \Ex{K | \req(V, K) = 1}$.

    \item The agent's per-round utility is $\nicefrac{\sum_j v_j}{\sum_j \ell_j}$ and the total fraction of rounds she holds the item for is $\nicefrac{\sum_j k_j}{\sum_j \ell_j}$. Since $(\ell_j, k_j, v_j)$ is independent across different epochs, as the number of epochs approaches infinity, we get that her expected per-round utility is $\nicefrac{\Ex{v_1}}{\Ex{\ell_1}}$ and the expected fraction of rounds she holds the item for is $\nicefrac{\Ex{k_1}}{\Ex{\ell_1}}$.
\end{itemize}

Using these facts, we can re-parameterize and re-write the optimization problem (\ref{eq:ideal:MDP}) as follows:
\begin{equation}
\label{eq:ideal:request}
\begin{split}
    \max_{\req} \qquad & \frac{\Ex{ V K \big| \req(V, K)=1 }}{\frac{1}{q} - 1 + \Ex{K \big| \req(V, K)=1}}  
    \\
    \textrm{such that} \qquad
    & \Pr{ \req(V,K)=1 } = q
    \\
    & \frac{\Ex{K \big| \req(V, K)=1}}{\frac{1}{q} - 1 + \Ex{K \big| \req(V, K)=1}} \le \a
\end{split}
\end{equation}

Before we show how an agent can efficiently solve the optimization problem (\ref{eq:ideal:request}), we make some observations about its optimal solution.
\begin{itemize}
    \item One natural question is whether the optimal request policy of the agent is independent of $K$ (and in particular, if $\req(V, K) = \One{V \ge \bar v}$ for some $\bar v$); note that this is the case in the settings with single-round demands.
    However, the following example shows this is not the case for reusable resources:
    Consider an agent with fair share $\a$ and the following distribution on $(V, K)$
    \begin{align*}
        (V, K) =
        \begin{cases}
            (1, 1), &\textrm{ with probability } \a/2 \\
            (\e, 2), &\textrm{ with probability } \a/2 \\
            (\e^2, 1), &\textrm{ otherwise}
        \end{cases}
    \end{align*}
    for some $\e$ much smaller than $\a$. It is easy to see that the optimal request policy should have $\req(1, 1) = 1$ with probability $1$. However, if $\req(V, K) = 0$ in the other cases, then the agent gets the item for only $\a/2$ fraction of the rounds. In order to increase her ideal utility the agent can set $\req(\e^2, 1) = 1$ with some probability, but in the optimal solution, it should always be $\req(\e, 2) = 0$. Intuitively, getting the item for $2$ rounds and gaining only $2\e$ utility on those rounds, hinders the agent from getting expected utility $\a/2$ on the next round (recall that $\a$ is much larger than $\e$). Formally, if $\req(\e, 2) = 1$ with positive probability, the denominator in the objective function of \cref{eq:ideal:request} becomes much larger, while the numerator increases only slightly, overall decreasing the ideal utility.

    \item The above example also shows that given fair share $\alpha$, it is possible that the optimal request policy by the agent results in resource-usage fraction less than $\a$ (we will need to distinguish this in our proofs later). In particular, if the agent's value is only $(1, 1)$ or $(\e, 2)$, then as we argue above, the optimal request policy sets $\req(\e, 2) = 0$ with probability $1$, resulting in the agent holding the item with probability $\a/2$.
\end{itemize}

Finally, in the case where $\Theta$ is a \emph{finite} type-space, we can convert the optimization problem \eqref{eq:ideal:request} into a linear program as shown in the lemma that follows. The lemma shows that in the case where $\Theta$ is a finite type set, then the optimal request policy $\req(\theta)$ underlying the ideal utility can be solved efficiently via a linear program. Subsequently, we will use this policy as a black box for defining the agent's robust bidding strategy in the pseudo-market.

In the linear program below, for each type $\type$, we use variables $f_\type$ to denote the expected fraction of rounds in which the resource is available, the agent has type $\type$, and she requests to reserve the resource (note that this is not the fraction of rounds the agent uses the resource while \textit{having} demand type $\theta$ -- this is $k_\theta f_\theta$). Using these variables, we can restrict the agent to use the resource in at most a fraction $\a$ rounds using a linear inequality. In addition, we need to bound each $f_\type$ to be at most as much as type $\type$ is available.

\begin{lemma}
\label{lem:ideal:LP}
    The optimization problem \eqref{eq:ideal:request} can be constructed as follows:
    Suppose that the agent has each type $\type=(v_\type, k_\type)\in\Theta$ with probability $p_\type$, and let $x_\type = p_\type \Pr{\req(\type) = 1 | \type}$ denote the probability that the agent has type $\theta$ and requests the item given its availability. Then we have $x_\type  = \frac{f_\type}{1 - \sum_{\type'} (k_{\type'} - 1) f_{\type'}},$ where $\{f_\type\}$ is the solution to the following linear program.
    \begin{equation}
    \label{eq:ideal:LP}
        \begin{split}
            \max_{\{f_\theta\}_{\theta\in\Theta}} \qquad
            & \sum_\type v_\type k_\type f_\type
            \\
            \textrm{such that} \qquad
            & \sum_\type k_\type f_\type \le \a
            \\
            & 0 \le f_\type \le p_\type \left( 1 - \sum_{\type'} (k_{\type'} - 1) f_{\type'} \right)\qquad\forall\,\theta\in\Theta
        \end{split}
    \end{equation}
\end{lemma}

To understand the conversion between $x_\type$ and $f_\type$ and the upper bound used for $f_\type$, note that when the agent gets the item for $k$ rounds at some time $t$ then on the following $k-1$ rounds the item is not available. This means that the fraction of rounds the item is available is $1 - \sum_{\type'} (k_{\type'} - 1) f_{\type'}$. On any of those rounds, the probability of having type $\theta$ is $p_\theta$, which results in the upper bound on $f_\theta$ given above.

\begin{myproof}[Proof of~\cref{lem:ideal:LP}]
    Using the variables $\{x_\type\}_{\type\in\Theta}$ we rewrite \cref{eq:ideal:request}:
    \begin{equation*}
    \begin{split}
        \max_{\{x_{\theta}\}_{\theta\in\Theta}} \qquad
        & \frac{\sum_\type v_\type k_\type x_\type}{1 + \sum_\type (k_\type - 1)x_\type}
        \\
        \textrm{such that} \qquad
        & \frac{\sum_\type k_\type x_\type}{1 + \sum_\type (k_\type - 1) x_\type} \le \a
        \\
        & 0 \le x_\type \le p_\type\qquad\qquad\qquad\forall\,\theta\in\Theta
    \end{split}
    \end{equation*}

    We turn the above into an LP by setting $f_\type = \frac{x_\type}{1 + \sum_{\type'} (k_{\type'} - 1) x_{\type'}}$, which is equivalent to the the linear system $\left(I - \vec f \left(\vec k - 1\right)^\top\right) \vec{x} = \vec f$. Now, as long as $1 - \vec f^\top \left(\vec k - 1\right) = 1 - \sum_{\type'} (k_{\type'} - 1) f_{\type'} \ne 0$ (which holds due to the constraints imposed on $\vec f$ as shown below), we can use the Sherman-Morrison matrix inversion formula\footnote{\url{https://en.wikipedia.org/wiki/Sherman-Morrison_formula}} to get $\left(I - \vec f \left(\vec k - 1\right)^\top\right)^{-1} = I + \frac{\vec f \left(\vec k - 1\right)^\top}{1 - \vec f^\top \left(\vec k - 1\right)}$. Thus, we get the unique solution $x_\type = \frac{f_\type}{1 - \sum_{\type'} (k_{\type'} - 1) f_{\type'}}$, and substituting this in the above program, we get the promised LP in~\cref{eq:ideal:LP}. 
\end{myproof}

\subsection{Ideal Utility and Social Welfare}

As we mentioned at the beginning of the section, the ideal utility provides a benchmark for how much utility each agent can hope to achieve, independent of other agents. This is important in our setting since the absence of money implies no way to make interpersonal comparisons between agents. However, in the special case when all agents are \emph{symmetric} (i.e., have the same type distributions), then even without knowing this distribution, we can directly reason about their relative values, and consequently, maximizing social welfare becomes a reasonable goal. 
In this setting, if all fair shares are set to be the same (and therefore, where every agent has the same ideal utility $\vstar$), $n \vstar T$ is an upper bound for the expected maximum social welfare that can be achieved if a central coordinator allocates the resource every round with knowledge of every agent's current demand but not future ones.
To see why this is true, note that the optimal coordinator would need to use a policy to allocate the item in round $t$ like the one in the \cref{def:ideal:single} but which depends on all the agents' current demands, is unconstrained, and chooses which agent (if any) to allocate to. Since the agents are symmetric we can assume that the optimal policy is also symmetric across agents. Because of this symmetry, every agent $i$ gets the item for at most an $\a_i = \nicefrac 1 n$ fraction of the rounds in expectation, making her per-round expected utility less than $\vstar$ (guaranteed by her optimal constrained policy of optimization problem \eqref{eq:ideal:MDP}), which implies the upper bound.

The above upper bound entails that every per-agent robustness guarantee for ideal utility also implies a social welfare guarantee. For example, our guarantee in \cref{thm:guar:guarantee} that every agent can realize at least half her ideal utility in expectation, implies a $\nicefrac 1 2$ approximation-ratio for the optimal social welfare that a central coordinator without future knowledge could achieve.
We emphasize, however, that in most cases $n \vstar T$ is a loose upper bound. For example, if there are $n$ agents, with single-round demands, and where each agent has value $1$ with probability $\nicefrac{1}{n}$, and $0$ otherwise, then with equal shares we have $\vstar = \nicefrac{1}{n}$, giving a bound of $n \vstar T = T$. However, since on average a $\nicefrac{1}{e}$ fraction of rounds have $0$ value for all agents, the social welfare is at most $(1-\nicefrac{1}{e})T$.
\section{Allocating a Single Reusable Resource: Robustness Guarantees}
\label{sec:guar}

Given the above setup, we are now ready to characterize the performance of the \mechanism~\cref{algo:algo}). In particular, our main result is the following \emph{per-agent robustness guarantee} that the mechanism enjoys: we show that under a reserve price $r$, \emph{every agent can get a constant fraction (depending on $r$) of their ideal utility, irrespective of how other agents behave}.
With respect to this robustness guarantee, we show that the minimax optimal reserve $r$ is $2$, in which case the agent can ensure they get at least half their ideal utility.

Before proceeding, we need to introduce some notation. Since we are studying guarantees from the perspective of a single agent, we henceforth drop the $i$ subscript.
Throughout this section, we define $\vstar$ to be the ideal utility of the agent and $\b$ to be the fraction of rounds in which the agent claims the item under the optimal request policy $\req$, when there is no competition. More specifically, from \cref{eq:ideal:request}, we have
\begin{equation}
\label{eq:vstar_beta}
\begin{split}
    \vstar &= \frac{\Ex{V K \big| \req(V, K) = 1}}{\frac{1}{q} - 1 + \Ex{K \big| \req(V, K) = 1}}
    \\
    \b &= \frac{\Ex{K \big| \req(V, K) = 1}}{\frac{1}{q} - 1 + \Ex{K \big| \req(V, K) = 1}} \le \a
\end{split}
\end{equation}
Finally, we assume there is some upper bound $\kmax$ on the duration of demands that agents sample (i.e., $K_{i}[t] \le \kmax$ for all $i, t$).

To prove our robustness bound, we consider the following simple bidding strategy for an agent: 
\begin{tcolorbox}[standard jigsaw,opacityback=0,label=def:policy]
    \textbf{\policy}:\\ 
    -- Agent solves the LP in \cref{lem:ideal:LP} to compute the request probability $\Pr{\req(\type) = 1 | \type}$ for each type $\theta$ that realizes the ideal utility for her fair share $\a$.\\ 
    -- For round $t$ the agent re-samples $\req(\type[t])$.\\
    -- If in round $t$ the item is available, the agent has enough budget ($B[t]\geq rK[t]$), and $\req(\type[t])=1$, then she competes in the auction with (per-round) bid $b[t]=r$ and duration $d[t]=K[t]$. 
\end{tcolorbox}

In other words, the agent computes her optimal stationary request policy in the no-competition setting, and then, while her budget is sufficient, bids at the reserve price according to this request policy. Note that this policy can be computed efficiently using~\cref{lem:ideal:LP}. Now, to understand the performance of this strategy, we first establish a simple lemma that shows that the agent's utility in each round under the \policy is directly proportional to her payment in that round.
For simplicity, we henceforth assume that if an agent who follows \policy wins the auction in round $t$ for $K[t]$ rounds, then she pays the total amount $P[t] = r K[t]$ and realizes her total utility $U[t] = V[t] K[t]$ instantaneously.

\begin{lemma}\label{lem:guar:BPB}
    Consider an agent with fair share $\a$, and compute her ideal utility $\vstar$ and ideal utilization fraction $\b\leq\a$.
    Let $\type[t]=(V[t], K[t])$ denote her demand type in round $t$, and let $(U[t], P[t])$ be her total realized utility and total payment in round $t$ under \policy (and any fixed policy of other agents).
    Then, for all $t\in[T]$, we have    
    \begin{align*}
        \Ex{U[t]} = \frac{\vstar}{\beta r} \Ex{P[t]}
    \end{align*}
\end{lemma}

This lemma utilizes the simplicity of the \policy, and in particular, the fact that the agent's utility is positive in round $t$ only if $P[t] = r K[t]$ and that the agent winning on round $t$ and $\type[t]$ are independent conditioned on $\req(\type[t]) = 1$.

\begin{myproof}
    Let $W[t]$ be an indicator random variable that is $1$ if the agent wins in round $t$, else $W[t] = 0$.
    We are going to use the fact that the agent's type $\type[t] = (V[t], K[t])$ does not directly depend on $W[t]$, but rather on whether the agent chooses to bid, which is determined by the request policy $\req(V[t], K[t])$ (formally, $\type[t]$ is conditionally independent of $W[t]$ given $\req(\type[t])$).
    Now we write
    \begin{alignat*}{3}
        \Line{
            \Ex{U[t]}
        }{=}{
            \Ex{K[t] V[t] W[t]}
        }{}
        \\
        \Line{}{=}
        {
            \Ex{K[t] V[t] \big| W[t]=1} \Pr{W[t]=1}
        }{}
        \\
        \Line{}{=}
        {
            \Ex{K[t] V[t] \big| \req(\type[t]) = 1} \Ex{W[t]}
        }{}
        \\
        \Line{}{=}
        {
            \frac{\vstar}{\frac{1}{q} - 1 + \Ex{K[t] \big| \req(\type[t])=1}}
            \Ex{W[t]}
        }{}
    \end{alignat*}
    where in the last equality we used \cref{eq:vstar_beta}. Similarly,
    \begin{alignat*}{3}
        \Line{
            \Ex{P[t]}
        }{=}{
            \Ex{r K[t] W[t]}
        }{}
        \\
        \Line{}{=}
        {
            r \Ex{K[t] \big| W[t]=1} \Pr{W[t]=1}
        }{}
        \\
        \Line{}{=}
        {
            r \Ex{K[t] \big| \req(\type[t])=1} \Ex{W[t]}
        }{}
        \\
        \Line{}{=}
        {
            \frac{r \b}{\frac{1}{q} - 1 + \Ex{K[t] \big| \req(\theta[t])=1}}
            \Ex{W[t]}
        }{}
    \end{alignat*}
    where in the last equality we used \cref{eq:vstar_beta}.
    Combining the two above equalities we get the desired bound.
\end{myproof}

We now proceed to prove the main result of this section, that using \policy, an agent can guarantee a fraction of her ideal utility.

\begin{theorem}\label{thm:guar:guarantee}
    Consider \mechanism with reserve price $r \ge 1$, and any agent with fair share $\a$, corresponding ideal utility $\vstar$, and ideal utilization fraction $\b$. Then, using \policy, irrespective of how other agents bid, the agent can guarantee a total expected utility
    \begin{align*}
        \Ex{\sum_{t\in[T]}U[t]}\geq \vstar T \left(\min\left\{ \frac{1}{r} , 1 - \frac{1}{r} \right\} - \frac{1}{\b} O\left( \frac{\kmax}{\sqrt{T}}\right)\right)
    \end{align*} 
\end{theorem}

\begin{remark}
    The first term in the competitive ratio in~\cref{thm:guar:guarantee} is maximized when $r = 2$, in which case it becomes $\nicefrac{1}{2}$. We note also that via more careful accounting, the first term can be improved to $\min\left\{ \frac{1}{r} , 1 - \frac{1-\a}{r} \right\}$, which gives the optimal result for $\a=1$; for ease of presentation, we defer this improvement to~\cref{sec:multi} (see~\cref{thm:mult:guarantee} and~\cref{sec:app:missing_proofs}).
\end{remark}

At a high level (and ignoring the sub-linear in $T$ terms), the proof of the theorem rests on a critical property of the reserve price that irrespective of the bidding policy, \emph{it limits the number of rounds that other agents can claim at prices greater than or equal to the reserve}. The rest of the agents have a budget of $(1-\a)T$, and hence at a reserve price of $r$, they can collectively win at most $(1-\a)T/r$ rounds at a bid that is higher than the reserve. More precisely, owing to the structure of the \mechanism, the other agents can collectively \emph{block} the agent on at most $(1-\a)T/r$ rounds. The remaining rounds are available to the agent, and roughly a $\b$ fraction of these have high value (i.e., are requested for under $\req$). 
In particular, if we choose $r=2$, then the other agents can block at most $T/2$ rounds, and among the remaining rounds, the agent wants roughly $\b T/2$ rounds and also has sufficient budget to claim these rounds at the reserve price. 

The problem with formalizing the above intuition is that with multi-round allocations, it is not enough to show that the agent has access to a $\b/2$ fraction of the rounds on average, as in order to receive utility for a demand $(V[t], K[t])$, the agent must get the item for the entire interval $[t,t+K[t]-1]$. As an extreme case, suppose $\a$ is small, and the agent had type either $(1,1)$ or $(0,0)$, and moreover, on a given sample path, had $(V[t], K[t])=(0,0)$ on almost all odd rounds. Now the remaining agents could `adversarially' bid $(2r,1)$ on all even rounds, and by blocking these, ensure the agent rarely wins the item. Of course, such a sample path is unlikely with i.i.d. demands, but the challenge still is to rule out all such bidding behavior by the other agents.

The main idea in our proof is to track all the rounds in which the agent is not \emph{blocked} -- i.e., where the item is available and all other agents bid lower than $r$ -- and argue that by playing the \policy, the agent wins close to a $q$ fraction of these rounds (recall $q$ is the probability that $\req(V[t], K[t]) = 1$ in the solution of \cref{eq:ideal:request}). Such a property is true for any set of rounds \emph{fixed upfront}; however, the set of non-blocked rounds may not be independent, both because they may be blocked adversarially, and also due to the lengths of the reservations. 
Additionally, because the agent's final payment is the minimum between her budget and a function of the unblocked rounds she has a high value for, in order to lower bound her payment we need a high probability bound on the second quantity.
We do that by showing that at any time $t$, the difference between the utilization of the agent up to $t$, and the number of unblocked rounds up to $t$ scaled by $\b$, forms a sub-martingale, which lets us use the Azuma-Hoeffding inequality\footnote{\url{https://en.wikipedia.org/wiki/Azuma's_inequality}} to get the high probability bound. We then combine this with the fact that the number of blocked rounds overall is at most $T/r$ to get the desired result.

\begin{myproof}[Proof of \cref{thm:guar:guarantee}]
    Fix our agent, Alice, with fair share $\a$, and corresponding ideal utility $\vstar$ and utilization $\b\leq \a$. We assume that Alice plays \policy: in any round $t$, if the item is available and $\req(V[t], K[t]) = 1$, then she participates in \mechanism with per-round bid $b[t]=r$ and duration $d[t]=K[t]$.

    Let $\Blk[t]\in\{0,1\}$ be an indicator variable that Alice is \emph{blocked} from competing for the item in round $t$. In particular, we have $\Blk[t] = 1$ if
    \begin{itemize}
        \item The item was reserved for round $t$ by Alice in a previous round (at a rate of $r$ credits per day for the item).
        \item The item was previously reserved for round $t$ by an agent other than Alice, who is paying at least $r$ per round.
        \item The item is available, but some agent other than Alice bids at least $r$ (for one or multiple rounds), and wins the round $t$ auction.
    \end{itemize}
    
    In all other cases, we have $\Blk[t] = 0$. 

    Given $\Blk[t]$, and assuming Alice has a remaining budget of at least $rK[t]\leq r\kmax$, we have that her payment $P[t]$ in round $t$ is
    \begin{align*}
        P[t] = r K[t] \One{\textrm{Alice wins auction in round } t}
        = r K[t] \req(\type[t]) (1 - \Blk[t])
    \end{align*}

    Since Alice might become budget limited at some point, her overall payment is at least 
    \begin{align}\label{eq:1:1}
        \sum_t P[t]
        \ge
        \min\left\{
            \a T, r \sum_{t=1}^T K[t] \req(\type[t]) (1 - \Blk[t])
        \right\} - r k_{\max}
    \end{align}
    We now study the second term in the minimum, which represents Alice's \emph{unconstrained} spending (i.e., if she was never budget limited), and prove a high probability lower bound on this. Subsequently, using~\cref{lem:guar:BPB}, we can translate this into a utility lower bound. To this end, we define
    \begin{align*}
        Z_\tau
        =
        \sum_{t=1}^\tau K[t] \req(\type[t]) (1 - \Blk[t])
        -
        \b\sum_{t=1}^\tau (1 - \Blk[t])
    \end{align*}

    To understand the rationale behind this, briefly assume that Alice makes requests that last only one round, i.e., $\req(\type[t]) = 1$ implies $K[t] = 1$. Given any set $\mathcal{T}$ of rounds chosen \emph{independently} of $\req$, if Alice wins \emph{all} the rounds in $\mathcal{T}$ she requests for under policy $\req$, then she would expect to utilize the item for at least $\b |\mathcal{T}|$ rounds in $\mathcal{T}$.
    Now observe that, over the first $\tau$ rounds, the set of rounds $\{t\leq \tau: \Blk[t]=0\}$ are precisely those on which Alice is not blocked from the item, and $Z_\tau$ counts the difference between the actual number rounds (budget-unconstrained) Alice wins in this set, and $\b$ times the number of rounds in the set. If the set was chosen independently of Alice's policy, then $Z_\tau$ would be a martingale, which we can then use to estimate the total expected payment $\Ex{\sum_t P[t]}$, and hence the total utility.

    Unfortunately, however, with no additional assumptions on the bidding behavior of other agents, we can not assert that the set of unblocked rounds is independent of the $\req$ policy (for example, the adversary's policy may depend on her budget). Nevertheless, we show below that $Z_\tau$ is a sub-martingale with respect to the history of the previous rounds $\mathcal H_{\tau-1}$. 
    
    First, recall we define $q = \Pr{\req(\type[t])=1}$ under Alice's optimal request policy (\cref{eq:ideal:request}). Moreover, we also have $q = \Pr{\req(\type[t])=1|H_{t-1}}$, since $\req$ is a \emph{stationary} policy that only depends on $\type[t]$, the agent's type in round $t$, which is independent of $\mathcal H_{t-1}$. Thus, we have:
    \begin{alignat*}{3}
        \Line{
            \Ex{Z_\tau - Z_{\tau - 1} \big| \mathcal H_{\tau-1}} 
        }{=}{
            \Ex{K[\tau] \req(\type[\tau]) (1 - \Blk[\tau]) \big| \mathcal H_{\tau-1}} - \b \Ex{(1 - \Blk[\tau]) \big| \mathcal H_{\tau-1}}
        }{}
        \\
        \Line{}{=}{
            q\Ex{K[\tau] (1 - \Blk[\tau]) \big| \req(\type[\tau])=1, \mathcal H_{\tau-1}} - \b \Ex{(1 - \Blk[\tau]) \big| \mathcal H_{\tau-1}}
        }{}
    \end{alignat*}
    Next, from~\cref{eq:vstar_beta}, we get that $\Ex{K[t] \big| \req(\theta[t]) = 1} = \frac{1-q}{q}\frac{\b}{1-\b}$. Moreover, note that $K[\tau]$ and $\Blk[\tau]$ are independent given that Alice wants to request the item in round $\tau$ (i.e., $K[\tau]$ and $\Blk[\tau]$ are conditionally independent given $\req(\theta[t])=1$). Substituting, in the above equation, we get
    \begin{alignat*}{3}
        \Line{
            \Ex{Z_\tau - Z_{\tau - 1} | \mathcal H_{\tau-1}} 
        }{=}{
            \frac{\b(1-q)}{(1-\b)}
            \Ex{(1 - \Blk[\tau]) \big| \req(\type[\tau]), \mathcal H_{\tau-1}} - \b \Ex{(1 - \Blk[\tau]) \big| \mathcal H_{\tau-1}}
        }{}
        \\
        \Line{}{\ge}{
            \b
            \Ex{(1 - \Blk[\tau]) \big| \mathcal H_{\tau-1}} - \b \Ex{(1 - \Blk[\tau]) \big| \mathcal H_{\tau-1}} = 0
        }{}
    \end{alignat*}
    where the second inequality follows from the facts that $\Blk[\tau]$ can only increase if we remove the condition that $\req(\type[\tau]) = 1$ and that $q \le \b$ (by \cref{eq:vstar_beta} and the fact that $K \ge 1$).

    Now, since $\Ex{Z_\tau - Z_{\tau - 1} | \mathcal H_{\tau-1}} \ge 0$, we have that $Z_\tau$ is a sub-martingale. Moreover, since $|Z_\tau - Z_{\tau-1}| \le \kmax$ with probability $1$, we can use Azuma's inequality to get for any $\e > 0$,
    \begin{align*}
        \Pr{Z_T - Z_0 \le  -\e}
        \le
        e^{-\frac{\e^2}{2T\kmax^2}}.
    \end{align*}
    In other words, for any $\e>0$, we have that with probability at least $1 - e^{-\frac{\e^2}{2T\kmax^2}}$
    \begin{align*}
    %\label{eq:1:2}
        r \sum_{t=1}^T K[t] \req(V[t], K[t]) (1 - \Blk[t])
        \ge
        r\b\sum_{t=1}^T (1 - \Blk[t]) - r \e
    \end{align*}

    On the other hand, note that we have $\sum_t \Blk[t] \le T \frac{1}{r}$ with probability $1$ -- this follows from the fact that for every round in which Alice is blocked, some agent (including possibly Alice) pays at least $r$ credits. Thus we have with probability at least $1 - e^{-\frac{\e^2}{2T\kmax^2}}$
    \begin{align}\label{eq:1:2}
        r \sum_{t=1}^T K[t] \req(V[t], K[t]) (1 - \Blk[t])
        \ge
        r\b T \left(1 - \frac{1}{r}\right) - r \e
    \end{align}
    Combining \cref{eq:1:1,eq:1:2}, we get with probability at least $1 - e^{-\frac{\e^2}{2T\kmax^2}}$, Alice's payment satisfies
    \begin{align*}
        \sum_t P[t]
        \ge 
        \min\left\{
            \a T, r\b T \left(1 - \frac{1}{r}\right) - r \e
        \right\} - r k_{\max}
        \ge
        \b T r \min\left\{ \frac{1}{r} , 1 - \frac{1}{r} \right\} - r \e - r k_{\max}
    \end{align*}
    Setting $\e=\Theta(\kmax\sqrt{T})$ and taking expectations, we get that her expected payment is at least
    \begin{align*}
        \sum_t \Ex{P[t]}
        \ge
        \b T r \min\left\{ \frac{1}{r} , 1 - \frac{1}{r} \right\} - r k_{\max} - r O\left( \kmax \sqrt{T} \right)
    \end{align*}
    Finally, using \cref{lem:guar:BPB}, we get that Alice's expected utility satisfies
    \begin{align*}
        \sum_t \Ex{U[t]}
        \ge
        \vstar T \min\left\{ \frac{1}{r} , 1 - \frac{1}{r} \right\} - \frac{\vstar}{\b}k_{\max} - \frac{\vstar}{\b} O\left( \kmax\sqrt{T} \right)
    \end{align*}
    Finally, since $T \ge 1$ we get our promised guarantee.
\end{myproof}

In \cref{thm:guar:guarantee} we proved that an agent can guarantee approximately a $\max\{1/r, 1-1/r\}$ fraction of her ideal utility, by using a very simple strategy that involves only bidding $r$ or $0$. It is reasonable to wonder if the agent can guarantee a bigger fraction by using a more complicated strategy or if the choice of $r = 2$ is the optimal one. In the next theorem, we prove that the answer is negative. More specifically, we prove that our result in \cref{thm:guar:guarantee} is asymptotically tight under our mechanism as we jointly scale $T$, $\a$, and $k_{\max}$ (in particular, for large $T$, and assuming $\kmax = \omega(1)$ and $\a = o(1)$ with respect to $T$).
%as $\kmax$ gets bigger and $\a$ gets smaller \gfcomment{not sure if this is the best way to say the last thing. Maybe we can say if $\kmax = \omega(1)$ and $\a = o(1)$}.\etcomment{what you propose as alternate seems better to me}

\begin{theorem}\label{thm:guar:impossibility}
    Consider the \mechanism mechanism, with maximum reservation duration $k_{\max} \ge 2$, and reserve price $r \ge 0$. Then there is a strategy for the other agents such that an agent with budget $\a T$ and ideal utility $\vstar$ is limited to
    \begin{align*}
        \Ex{\sum_{t\in[T]}U[t]}\leq \vstar T \left(
            1 - \frac{(1-\a)}{\max\{1,r\}}
            + \frac{1}{k_{\max}}
        \right)
        +
        \vstar(\kmax - 1)
        % + \frac{\vstar}{\b}\sqrt{T \log(1/\d)/2}
    \end{align*}
    % 
    % with probability at least $1-\d$, for any $\d \ge 0$.
\end{theorem}

In order to get the bound we consider an agent with a (small) fair share $\a$, and moreover, who in each round has demand $\type[t]=(1,1)$ with probability $\a$, and $(0,0)$ otherwise.
In contrast, suppose every time the item is available, at least one other agent demands $\kmax$ rounds at price $\max\{1,r\}$. 
This means that the agent has three options to get utility.
First, she can bid slightly more than $\max\{r,1\}$ only when she has positive value and get the item if it is available (which only happens with probability $1/\kmax$) until all other agents run out of money. 
Second, she can bid slightly more than $\max\{r,1\}$ when she has zero value to stop the other agents from getting and blocking the item.
Third, she can wait for the other agents' budget to deplete, which happens in the last $T(1 - \frac{1-\a}{\max\{1,r\}})$ rounds.

\begin{myproof}
    Fix our agent, her budget $\a T$, and her type to be $(V[t], K[t]) = (1, 1)$ with probability $\a$ and $(V[t], K[t]) = (0, 1)$ otherwise. We notice that $\vstar = \a$ and $\b = \a$.

    We will show the impossibility result by assuming that all the other agents follow the same deterministic strategy: each tries to reserve $k_{\max}$ rounds, by bidding $\max\{1,r\}$ per round. For the rest of the proof, we are going to think of all the other agents into one adversary, with a total budget of $(1-\a)T$.
    
    Let $A[t] \in \{0,1\}$ denote the availability of the item: if $A[t] = 1$ then in round $t$ the agent has the ability to bid and reserve the item; if $A[t] = 0$ then the adversary has reserved the item in a previous round and still has it. As long as $A[t] = 1$, the agent can control if she wins the item in round $t$ or not by bidding slightly above $\max\{1,r\}$. This allows us to assume w.l.o.g. that she only requests the item for $1$ round at a time since requesting the item for multiple rounds can be simulated by requesting the item for $1$ round multiple times. We notice that the agent's utility is
    \begin{align*}
        \sum_{t=1}^T U[t]
        =
        \sum_{t=1}^T V[t] \One{ \textrm{agent wins in } t }
        =
        \sum_{t=1}^T A[t] V[t] \One{ \textrm{agent bids in } t }
        \le
        \sum_{t=1}^T A[t] V[t].
    \end{align*}

    We now notice that since the random variables $A[t]$ and $V[t]$ are independent, $\Ex{A[t] V[t]} = \a\Ex{A[t]}$. This proves that the agent's expected utility
    \begin{align}\label{eq:1:3}
        \Ex{\sum_{t=1}^T U[t]}
        \le
        \a\Ex{\sum_{t=1}^T A[t]}.
    \end{align}

    We now upper bound the sum of the above quantity. Let $U$ be the number of rounds the adversary wins, which makes the item unavailable for $(k_{\max}-1)U$ rounds in total. Note that $\sum_{t=1}^T A[t] = T - (k_{\max}-1) U$. We now lower bound $k_{\max} U$, the number of rounds the adversary holds the item, by observing that the adversary must eventually run out of budget (since as long as the adversary has budget either she or the agent pays at least $1$ for the item each round). This means that $k_{\max} U \ge \frac{(1-\a)T}{\max\{1,r\}} - k_{\max}$, i.e., the adversary gets the item until their budget is depleted, up to an additive error of $k_{\max}$. Combining this with our previous bound  for the agent's utility in \eqref{eq:1:3}, we have that
    \begin{align*}
        \Ex{\sum_{t=1}^T U[t]}
        \le
        \a\left(T - \frac{k_{\max}-1}{k_{\max}} \frac{(1-\a)T}{\max\{1,r\}} + \kmax - 1\right) 
    \end{align*}
    By rearranging the above, we complete the proof.
\end{myproof} 
\section{Impossibility Result for Single Resource Allocation}
\label{sec:hardness}

In this section, we study upper bounds on the fraction of ideal utility an agent can get under \emph{any} mechanism, including  mechanisms without a pseudo-market structure.
More specifically, we show that our result in \cref{thm:guar:guarantee} is tight: as the number of agents $n$ and the maximum number of days an agent can have demand for, $\kmax$, become large, no mechanism can guarantee every agent more than half their ideal utility.

\begin{theorem}\label{thm:imposs}
    There exists a sequence of settings with $n$ symmetric agents, each with fair share $\nicefrac{1}{n}$, and consequently the same ideal utility $\vstar$, but where, as $n\to\infty$, no mechanism can guarantee every agent total expected utility more than
    \begin{align*}
        \vstar T \left( \frac{1}{2} + O\left( \frac{1}{\kmax} \right) \right)
    \end{align*}
    % 
%    as $n\to\infty$.
\end{theorem}

We study an instance where the $n$ agents have zero value with high probability and positive values have duration $\kmax$. In our example, if every agent gets allocated the item every time she has a positive value, then her total expected utility is exactly $\vstar T$. However, because the demands of the agents overlap, no mechanism can guarantee such an allocation for every agent and in fact can guarantee at most half of that in expectation.

\begin{myproof}
    Consider an example with $n$ agents, each with budget $T/n$ and the same type distribution
    \begin{align*}
        (V, K) =
        \begin{cases}
            (1, \kmax) , & \textrm{ with probability } \frac{1}{\kmax(n-1) + 1} := p
            \\
            (0, 1), & \textrm{ otherwise } 
        \end{cases}
        .
    \end{align*}

    Note that given the above probability $p$, if an agent sets $\req(1, \kmax) = 1$ with probability $1$, then her ideal utilization is exactly $\b = \a = \nicefrac{1}{n}$. This means that
    the ideal utility of every agent is $\vstar = \nicefrac{1}{n}$. We now calculate the expected maximum social welfare in this setting. Let $p' = 1 - (1-p)^n$ be the probability that any agent has positive value in a certain round when the item is free. The expected number of rounds before an item is allocated in this mechanism is $\nicefrac{1}{p'} - 1$ and then the item is allocated for $\kmax$ rounds. This means that the fraction of rounds the item is allocated for is
    \begin{align*}
        \frac{\kmax}{\kmax - 1 + \frac{1}{p'}}
        % \overset{(n,k)\to(\infty,\infty)}{\longrightarrow}
        % \frac{1}{2}
    \end{align*}
    
    We show that the above fraction equals $\nicefrac{1}{2}$ as $n$ and $\kmax$ approach infinity. We notice that if $n$ is large enough, for any value of $\kmax$
    \begin{align*}
        p'
        =
        1 - (1-p)^n
        =
        1 - \left( 1 - \frac{1}{\kmax(n-1) + 1} \right)^n
        \approx
        1 - e^{-1/\kmax}
        \le
        \frac{1}{\kmax}
    \end{align*}
    which makes
    \begin{align*}
        \frac{\kmax}{\kmax - 1 + \frac{1}{p'}}
        \lessapprox
        \frac{\kmax}{\kmax - 1 + \kmax}
        =
        \frac{1}{2} + O\left( \frac{1}{\kmax} \right)
    \end{align*}

    This implies that the expected maximum social welfare is $T(\nicefrac{1}{2} + O(\nicefrac{1}{\kmax}))$. This means that under any mechanism, at least one agent is going to have expected utility at most $\nicefrac{1}{n}$ times that, which proves the theorem.
\end{myproof}
\section{Generalized Reusable Public Resource Allocation}
\label{sec:multi}

In this section, we extend our results to the case where there are $L \ge 1$ identical items shared among the agents. We still assume that each agent has nothing to gain by having more than one item on a single round. Simply adapting our single-resource result to multiple resources would result (ignoring lower order terms) in guaranteeing a 
$$\min\left\{ \frac{1}{r} , 1 - \frac{1 + \a (L-1)}{r} \right\}
$$
fraction of the agent's ideal utility (formally defined below). In his section, we show an improved guarantee of 
$$\min\left\{ \frac{1}{r} , 1 - \frac{1 - \a}{r} \right\}
$$
fraction, eliminating the deterioration of the guarantee as $L$ gets larger. In fact, the improved bound would also slightly improve our result in Section \ref{sec:guar}. There we focused on the simpler proof as the improvement in the case of $L=1$ is not so significant.

\subsection{Mechanism and Ideal Utility}

Our mechanism, \mechanism, is similar to the single resource case, except that now if in round $t$ there are $m$ available items, the agents with the top $m$ valid bids get allocated an item each. Additionally, in this case, we normalize the budgets such that the total budget is $L T$ and each agent $i$ has budget $\a_i L T$.

Our definition of ideal utility for agent $i$, in this case, is almost identical to the one in \cref{def:ideal:single}, except we allow each agent to request the item for at most $\a_i L$ fraction of the rounds, assuming $\a_i L\le 1$ (otherwise the ideal utility of the agent allows him to request the item every round). Specifically, the analogous LP of \cref{eq:ideal:request} that defines the agent $i$'s ideal utility when she has fair share $\a_i$ and $(V, K) \sim \mathcal F_i$ is
\begin{equation}\label{eq:mult:request}
\begin{split}
    \vstar = \max{}_\req \qquad
    & \frac{\Ex{ V K \big| \req(V, K) }}{\frac{1}{q} - 1 + \Ex{K \big| \req(V, K) }}
    \\
    \textrm{such that} \qquad
    & \Pr{ \req(V, K) } = q
    \\
    & \frac{\Ex{K \big| \req(V, K) }}{\frac{1}{q} - 1 + \Ex{K \big| \req(V, K) }} 
    = \b \le \a_i L
\end{split}
\end{equation}

\subsection{Guarantees of \texorpdfstring{\mechanism}{First-Price Pseudo-Auction with Multi-Round Reserves} for multiple resources}

For the rest of the section, we focus on guarantees that an agent can get when using our mechanism. Henceforth, we drop the $i$ subscript.

We are going to focus on the case when $\a L$ is small. We do so because this is the more interesting setting. Otherwise, very simple mechanisms like round-robin could make strong guarantees.

We proceed to prove that even in this more complicated setting, an agent who follows \policy, exactly as described in \cref{sec:guar}, can guarantee an almost $1/2$ fraction of their ideal utility in expectation. We start with a lemma that is identical to \cref{lem:guar:BPB}. We defer the proofs of this section to \cref{sec:app:missing_proofs}.

\begin{lemma}\label{lem:mult:BPB}
    Consider an agent with budget $\a L T$, and compute her ideal utility $\vstar$ and optimal utilization fraction $\b\leq\a L$.
    Let $\type[t]=(V[t], K[t])$ denote her demand type in round $t$, and let $(U[t], P[t])$ be her total realized utility and total payment in round $t$ under \policy (and any fixed policy of other agents).
    Then, for all $t\in[T]$, we have
    \begin{align*}
        \Ex{U[t]} = \frac{\vstar}{\beta r} \Ex{P[t]}
    \end{align*}
\end{lemma}

We now proceed to show the main result of the section, that any agent can guarantee a fraction of her ideal utility in expectation.

\begin{theorem}\label{thm:mult:guarantee}
    Consider \mechanism with reserve price $r \ge 1$, and any agent with budget $\a L T$, corresponding ideal utility $\vstar$, and optimal utilization fraction $\b$. Then, using \policy, irrespective of how other agents bid, the agent can guarantee a total expected utility
    \begin{align*}
        \Ex{\sum_{t\in[T]}U[t]}
        \geq
        \vstar T \left(\min\left\{ \frac{1}{r} , 1 - \frac{1 - \a}{r} \right\} - \frac{1}{\b} O\left( \frac{\kmax}{\sqrt{T}}\right)\right)
    \end{align*} 
\end{theorem}

\begin{remark}
    The term in \cref{thm:mult:guarantee} is maximized when $r = 2 - \a$, in which case it becomes $\frac{1}{2 - \a}$.
\end{remark}

We defer the proof of this theorem to the Appendix. We note that even though the result is quite similar to the one in \cref{thm:guar:guarantee} the proof for it requires more careful analysis. More specifically, as mentioned above, if we follow the steps of the proof in \cref{thm:guar:guarantee} we would get a $\min\{\nicefrac{1}{r}, 1 - \nicefrac{1 + \a(L-1)}{r}\}$ bound instead. In order to improve the bound we need to analyze carefully the rounds when the item is not available because the agent still holds it from a previous round. Previously we upper bounded the number of those rounds by $T \frac{\a}{r}$ which in this case would become $T \frac{\a L}{r}$ which in turn leads to the worse bound.

\printbibliography

\appendix
\section{Missing proofs from Section \ref{sec:multi}}
\label{sec:app:missing_proofs}

In this section, we include the missing proofs from \cref{sec:multi}.
The proof of \cref{lem:mult:BPB} is identical to the proof of \cref{lem:guar:BPB}; we include it here for completeness
\begin{myproof}[Proof of \cref{lem:mult:BPB}]
    Let $W[t]$ be an indicator random variable that is $1$ if the agent wins in round $t$, else $W[t] = 0$.
    We are going to use the fact that the agent's type $\type[t] = (V[t], K[t])$ does not directly depend on $W[t]$, but rather on whether the agent chooses to bid, which is determined by the request policy $\req(V[t], K[t])$ (formally, $\type[t]$ is conditionally independent of $W[t]$ given $\req(\type[t])$).
    Now we write
    \begin{alignat*}{3}
        \Line{
            \Ex{U[t]}
        }{=}{
            \Ex{K[t] V[t] W[t]}
        }{}
        \\
        \Line{}{=}
        {
            \Ex{K[t] V[t] \big| W[t]} \Ex{W[t]}
        }{}
        \\
        \Line{}{=}
        {
            \Ex{K[t] V[t] \big| \req(\type[t])} \Ex{W[t]}
        }{}
        \\
        \Line{}{=}
        {
            \frac{\vstar}{\frac{1}{q} - 1 + \Ex{K[t] \big| \req(V[t], K[t])}}
            \Ex{W[t]}
        }{}
    \end{alignat*}
    where in the last equality we used \cref{eq:mult:request}. Similarly,
    \begin{alignat*}{3}
        \Line{
            \Ex{P[t]}
        }{=}{
            \Ex{r K[t] W[t]}
        }{}
        \\
        \Line{}{=}
        {
            r \Ex{K[t] \big| W[t]} \Ex{W[t]}
        }{}
        \\
        \Line{}{=}
        {
            r \Ex{K[t] \big| \req(\type[t])} \Ex{W[t]}
        }{}
        \\
        \Line{}{=}
        {
            \frac{r \b}{\frac{1}{q} - 1 + \Ex{K[t] \big| \req(V[t], K[t])}}
            \Ex{W[t]}
        }{}
    \end{alignat*}
    where in the last equality we used \cref{eq:mult:request}.
    Combining the two above equalities we get the desired bound.
\end{myproof}

We now prove the main result of \cref{sec:multi}, \cref{thm:mult:guarantee}.

\begin{myproof}[Proof of \cref{thm:mult:guarantee}]
    The start of the proof is analogous to the proof of \cref{thm:mult:guarantee}. We follow the same outline till the use of the Azuma-Hoeffding inequality. To get the improved bound, we need to do more careful accounting of the rounds the agent cannot bid on the resource either if she already has one or if all resources are blocked by others.
    
    Fix our agent, Alice, with budget $\a L T$, and corresponding ideal utility $\vstar$ and utilization $\b\leq \a L$. We assume that Alice plays \policy: in any round $t$, if an item is available and $\req(V[t], K[t]) = 1$, then she participates in \mechanism with per-round bid $b[t]=r$ and desired duration $d[t]=K[t]$.

    Let $W[t]\in\{0,1\}$ be an indicator variable that indicates that Alice wins the auction in round $t$ while following our policy but having an unlimited budget. Using this notation her total payment (with limited budget) is
    \begin{align}\label{eq:mult:1}
        \sum_t P[t]
        \ge
        \min\left\{
            \a L T, r \sum_{t=1}^T K[t] W[t]
        \right\} - r k_{\max}
    \end{align}
    
    Let $\Blk[t]\in\{0,1\}$ be an indicator variable that Alice is \emph{blocked} from competing for the item in round $t$. In particular, we have $\Blk[t] = 1$ if
    \begin{itemize}
        \item An item was reserved for round $t$ by Alice in a previous round (at a rate of $r$ credits per round for the item).
        \item \textit{All} the items were previously reserved for round $t$ by agents other than Alice, who are paying at least $r$ per round.
        \item Some items are available, but agents other than Alice bid at least $r$ (for one or multiple rounds) for them, and win all the auctions in round $t$.
    \end{itemize}
    
    In all other cases, we have $\Blk[t] = 0$. 

    Let $\kappa = \Ex{K[t] | \req[t]}$ and recall that $\Ex{K[t] | \req[t]} = \frac{(1-q)\b}{q(1-\b)}$ and $q = \Pr{\req(\type[t])=1}$. Now we define
    \begin{align*}
        Z_\tau
        =
        \sum_{t=1}^\tau K[t] W[t]
        -
        q \kappa\sum_{t=1}^\tau (1 - \Blk[t])
    \end{align*}

    We show below that $Z_\tau$ is a sub-martingale with respect to the history of the previous rounds $\mathcal H_{\tau-1}$. Note that $q = \Pr{\req(\type[t])=1|\mathcal H_{t-1}}$, since $\req$ is a \emph{stationary} policy that only depends on $\type[t]$, which is independent of $\mathcal H_{t-1}$. Thus, we have:
    \begin{alignat*}{3}
        \Line{
            \Ex{Z_\tau - Z_{\tau - 1} \big| \mathcal H_{\tau-1}} 
        }{=}{
            \Ex{K[\tau] W[t] \big| \mathcal H_{\tau-1}} - q \kappa \Ex{(1 - \Blk[\tau]) \big| \mathcal H_{\tau-1}}
        }{}
        \\
        \Line{}{=}{
            \Ex{K[\tau] \req(\type[\tau]) (1 - \Blk[\tau]) \big| \mathcal H_{\tau-1}} - q \kappa \Ex{(1 - \Blk[\tau]) \big| \mathcal H_{\tau-1}}
        }{}
        \\
        \Line{}{=}{
            q\Ex{K[\tau] (1 - \Blk[\tau]) \big| \req(\type[\tau])=1, \mathcal H_{\tau-1}} - q \kappa \Ex{(1 - \Blk[\tau]) \big| \mathcal H_{\tau-1}}
        }{}
    \end{alignat*}
    Note that $K[\tau]$ and $\Blk[\tau]$ are independent given that Alice wants to request the item in round $\tau$ (i.e., $K[\tau]$ and $\Blk[\tau]$ are conditionally independent given $\req(\theta[\tau])=1$). Substituting, in the above equation, we get
    \begin{alignat*}{3}
        \Line{
            \Ex{Z_\tau - Z_{\tau - 1} | \mathcal H_{\tau-1}} 
        }{=}{
            q \kappa
            \Ex{(1 - \Blk[\tau]) \big| \req(\type[\tau]), \mathcal H_{\tau-1}} - q \kappa \Ex{(1 - \Blk[\tau]) \big| \mathcal H_{\tau-1}}
        }{}
        \\
        \Line{}{\ge}{
            q \kappa \Ex{(1 - \Blk[\tau]) \big| \mathcal H_{\tau-1}} - q \kappa \Ex{(1 - \Blk[\tau]) \big| \mathcal H_{\tau-1}} = 0
        }{}
    \end{alignat*}
    where the second inequality follows from the fact that $\Blk[\tau]$ can only increase if we remove the condition on $\req(\type[\tau]) = 1$.
    
    Now, since $\Ex{Z_\tau - Z_{\tau - 1} | \mathcal H_{\tau-1}} \ge 0$, we have that $Z_\tau$ is a sub-martingale. Moreover, since with probability $1$ it holds that $|Z_\tau - Z_{\tau-1}| \le \kmax$, we can use the Azuma-Hoeffding inequality to get for any $\e_1 > 0$,
    \begin{align*}
        \Pr{Z_T - Z_0 \le - \e_1}
        \le
        e^{-\frac{\e_1^2}{2T\kmax^2}}.
    \end{align*}
    In other words, for any $\e_1 > 0$, we have that with probability at least $1 - e^{-\frac{\e_1^2}{2T\kmax^2}}$
    \begin{align*}
        \sum_{t=1}^T K[t] W[t]
        \ge
        q \kappa\sum_{t=1}^T (1 - \Blk[t]) - \e_1
    \end{align*}

    For the proof of \cref{thm:guar:guarantee} this inequality was strong enough to yield the desired bound. To get the stronger bound we claimed here, we now `split' the indicator variable $\Blk[t]$ into two disjoint variables: $\Blk[t] = \Blk_1[t] + \Blk_2[t]$. $\Blk_1[t] = 1$ if another agent blocked Alice at round $t$ and $\Blk_2[t] = 1$ if Alice having the item in round $t$ because she won it in a previous round. Note that with these definitions $\Blk[t] = \Blk_1[t] + \Blk_2[t]$. We notice that because of the reserve price it holds that $\sum_{t=1}^T \Blk_1[t] \le T \frac{1-\a}{r}$. Additionally, because whenever Alice wins in round $t$ she blocks herself for the next $K[t] - 1 $ rounds, it holds that
    \begin{align*}
        \sum_{t=1}^T \Blk_2[t]
        \le
        \sum_{t=1}^T (K[t] - 1)W[t]
    \end{align*}

    Combining these facts with the bound above, we get that for any $\e_1 > 0$, we have that with probability at least $1 - e^{-\frac{\e_1^2}{2T\kmax}}$
    \begin{align}\label{eq:mult:21}
        \sum_{t=1}^T \big( (q \kappa + 1)K[t] - q \kappa \big) W[t]
        \ge
        q \kappa T \left( 1 - \frac{1-\a}{r} \right) - \e_1
    \end{align}

    The above bound does not seem particularly useful, because we are interested in $\sum_t K[t] W[t]$. We correlate the two quantities by using another martingale argument. We define
    \begin{align*}
        M_\tau
        =
        \sum_{t=1}^\tau \big( (q \kappa + 1)K[t] - q \kappa \big) W[t]
        -
        \big(1 + (\kappa - 1) q \big) \sum_{t=1}^\tau K[t] W[t]
    \end{align*}

    We are going to show that $M_\tau$ is a martingale with respect to the history of the previous rounds, $\mathcal H_{\tau-1}$. We utilize the fact that $W[t]$ and $K[t]$ are conditionally independent on $\req[t]$. More specifically, we have that (we omit the condition on $\mathcal H_{\tau-1}$ for brevity)
    \begin{alignat*}{4}
        \Line{
            \Ex{\big( (q \kappa + 1)K[\tau] - q \kappa \big) W[\tau]}
        }{=}{
            q \Ex{\big( (q \kappa + 1)K[\tau] - q \kappa \big) W[\tau] \big| \req(\type[t])}
        }{}
        \\
        \Line{}{=}{
            q
            \Ex{W[\tau] \big| \req(\type[t])}
            \Ex{\big( (q \kappa + 1)K[\tau] - q \kappa \big) \big| \req(\type[t])}
        }{}
        \\
        \Line{}{=}{
            q
            \Ex{W[\tau] \big| \req(\type[t])} \kappa
            \frac{\Ex{\big( (q \kappa + 1)K[\tau] - q \kappa \big) \big| \req(\type[t])}}{\kappa}
        }{}
        \\
        \Line{}{=}{
            q
            \Ex{W[\tau] K[\tau] \big| \req(\type[t])}
            \frac{\Ex{\big( (q \kappa + 1)K[\tau] - q \kappa \big) \big| \req(\type[t])}}{\kappa}
        }{}
        \\
        \Line{}{=}{
            \Ex{W[\tau] K[\tau]}
            \frac{ (q \kappa + 1)\kappa - q \kappa }{\kappa}
        }{}
    \end{alignat*}

    The above proves that $\Ex{M_\tau - M_{\tau-1} | \mathcal H_{\tau-1}} = 0$, i.e., $M_\tau$ is a martingale. Because it holds almost surely that $|M_\tau - M_{\tau-1}| \le \kmax$ we use Azuma-Hoeffding inequality and have that for any $\e_2 > 0$, with probability at least $1 - e^{-\frac{\e_2^2}{2T\kmax^2}}$
    \begin{align*}
        (1 + (\kappa - 1) q) \sum_{t=1}^\tau K[t] W[t]
        \ge
        \sum_{t=1}^\tau \big( (q \kappa + 1)K[t] - q \kappa \big) W[t]
        -
        \e_2
    \end{align*}

    Combining the above bound with \cref{eq:mult:21} and using the union bound we get that with probability at least $1 - e^{-\frac{\e^2}{2T\kmax^2}} - e^{-\frac{\e'^2}{2T\kmax^2}}$
    \begin{align*}
        \sum_{t=1}^\tau K[t] W[t]
        \ge
        \frac{q \kappa}{1 + (\kappa - 1) q} T \left( 1 - \frac{1-\a}{r} \right)
        - \frac{\e + \e'}{1 + (\kappa - 1) q}
    \end{align*}

    Substituting $\kappa = \frac{(1-q)\b}{q(1-\b)}$ and that $1 + (\kappa - 1) q \ge 1$ the above becomes
    \begin{align*}
        \sum_{t=1}^\tau K[t] W[t]
        \ge
        \b T \left( 1 - \frac{1-\a}{r} \right)
        - (\e + \e')
    \end{align*}

    Combining the above with \cref{eq:mult:1} and the fact that $\a L \ge \b$, we get with probability at least $1 - e^{-\frac{\e^2}{2T\kmax^2}} - e^{-\frac{\e'^2}{2T\kmax^2}}$, Alice's payment satisfies
    \begin{align*}
        \sum_t P[t]
        \ge
        \b T r \min\left\{ \frac{1}{r} , 1 - \frac{1-\a}{r} \right\} - r (\e + \e') - r k_{\max}
    \end{align*}
    Setting $\e = \e' = \Theta(\kmax\sqrt{T})$ and taking expectations, we get that her expected payment is at least
    \begin{align*}
        \sum_t \Ex{P[t]}
        \ge
        \b T r \min\left\{ \frac{1}{r} , 1 - \frac{1 - \a}{r} \right\} - r k_{\max} - r O\left( \kmax\sqrt{T} \right)
    \end{align*}
    Finally, using \cref{lem:mult:BPB}, we get that Alice's expected utility satisfies
    \begin{align*}
        \sum_t \Ex{U[t]}
        \ge
        \vstar T \min\left\{ \frac{1}{r} , 1 - \frac{1 - \a}{r} \right\} - \frac{\vstar}{\b}k_{\max} - \frac{\vstar}{\b} O\left( \kmax\sqrt{T} \right)
    \end{align*}
    which gets our promised guarantee.
\end{myproof}

\end{document}